\documentclass[11pt,a4paper]{article}
\pdfoutput=1
\usepackage{cancel}
\usepackage{enumitem}
\usepackage{etex}
\usepackage{amsthm}
\usepackage{geometry}
\usepackage{dcolumn}
\usepackage{epsf}
\usepackage{mathrsfs}
\usepackage{multirow}
\usepackage{booktabs}
\usepackage{tabularx}
\usepackage{array}
\usepackage{slashed}
\usepackage{float}
\usepackage{leftidx}
\usepackage{setspace}
\usepackage{verbatim}
\usepackage{adjustbox}
\usepackage{tikz}
\usepackage[caption=false]{subfig}
\usepackage{arydshln}
\usepackage{jheppub}
\usepackage{shuffle}
\usepackage{amsmath}
\usepackage{cases}

\numberwithin{equation}{section}
\usetikzlibrary{arrows,decorations.markings,shapes.arrows,patterns,positioning}

\newcommand{\bea}{\begin{eqnarray}}
\newcommand{\eea}{\end{eqnarray}}
\newcommand{\bean}{\begin{eqnarray*}}
\newcommand{\eean}{\end{eqnarray*}}
\newcommand{\nn}{\nonumber\\}
\newcommand{\Sl}{\sum\limits}

\def\W #1{\widetilde{#1}}

\def\Label#1{\label{#1}%
  \smash{\hbox to0pt{\raise1ex\hbox{\tiny[#1]}\hss}}}
\def\Label#1{\label{#1}}
\renewcommand{\eqref}[1]{eq.~(\ref{#1})}
\newcommand{\figref}[1]{Fig.~\ref{#1}}

\newcommand{\secref}[1]{section~\ref{#1}}

\def\Sl{\sum\limits}

\newcommand{\ctobedelete}[1]{}
\allowdisplaybreaks

\title{Note on graph-based BCJ relation for Berends-Giele currents}

\author[a,b,c]{Yi-Jian Du} \author[a]{Konglong Wu}

\affiliation[a]{School of Physics and Technology,
Wuhan University, \\
No.299 Bayi Road, Wuhan 430072, China}
\affiliation[b]{ Hubei Key Laboratory of Nuclear Solid Physics, Wuhan University,\\
No.299 Bayi Road, Wuhan 430072, China}
\affiliation[c]{Suzhou Institute of Wuhan University,\\
No.377 Linquan Street, Suzhou, 215123, China}

\emailAdd{yijian.du@whu.edu.cn, wukonglong@whu.edu.cn}

\date{\today}
\abstract{Graph-based Bern-Carasso-Johansson (BCJ) relation for Berends-Giele currents in bi-adjoint scalar (BS) theory, which is characterized by connected tree graphs, was proposed in an earlier work.  In this note, we provide a systematic study of this relation. We first prove the relations based on two special types of graphs: simple chains and star graphs. The general graph-based BCJ relation established by an arbitrary tree graph is further proved, through Berends-Giele recursion. When combined with proper off-shell extended numerators, this relation induces the graph-based BCJ relation for Berends-Giele currents in Yang-Mills theory. The corresponding relations for amplitudes are obtained via on-shell limits.

}
\keywords{ Amplitude Relations, Gauge invariance}

\begin{document}
\maketitle \flushbottom

\section{Introduction}





The expansion of tree-level Einstein-Yang-Mills (EYM) amplitudes has been studied in recent years \cite{Stieberger:2016lng,Nandan:2016pya,Schlotterer:2016cxa,Fu:2017uzt,Chiodaroli:2017ngp,Teng:2017tbo,Du:2017gnh}. Along the line of \cite{Fu:2017uzt,Du:2017gnh}, general recursive expansion formula, which expresses EYM amplitudes by those with fewer gravitons and (or) gluon traces, can be used to expand any gravity or EYM amplitude in terms of color-ordered Yang-Mills (YM) amplitudes. As shown in \cite{Fu:2017uzt,Du:2017kpo,Du:2017gnh} the expansion coefficients correspond to local Bern-Carrasco-Johansson (BCJ) numerators \cite{Bern:2008qj,Bern:2010ue} in Del Duca-Dixon-Maltoni (DDM) basis \cite{DelDuca:1999rs}. Systematic graphic rule has been proposed \cite{Du:2017kpo,Du:2017gnh,Hou:2018bwm,Du:2019vzf} to construct these coefficients and has been applied in many situations such as studying the relationship between the symmetry induced identities \cite{Hou:2018bwm,Du:2019vzf} and BCJ relations \cite{Bern:2008qj}, evaluating amplitudes in four-dimensions \cite{Tian:2021dzf}.

Since the expansion of gravity (GR) amplitudes in terms of YM amplitudes shares the same coefficients with the expansion of color-ordered YM amplitudes in terms of bi-adjoint scalar (BS) ones, one can also find the BCJ numerators by expanding  color-ordered YM amplitudes  \cite{Bern:2010yg,BjerrumBohr:2010hn, Du:2011js,Broedel:2013tta,Cachazo:2013gna,Cachazo:2013hca,Cachazo:2013iea,Cachazo:2014nsa,Mafra:2016ltu}. This expansion could be achieved in an off-shell way, by introducing expansion \cite{Mafra:2015syt,Lee:2016tbd,Bridges:2019siz,Frost:2020bmk,Wu:2021bcy} of Berends-Giele currents \cite{Berends:1987me}.  Such off-shell expansion helps one to understand amplitudes
 and in fact induces an off-shell construction of BCJ numerators. A key observation in the construction \cite{Wu:2021bcy} is that the Berends-Giele currents in BS satisfy a so-called graph-based BCJ relation, whose on-shell version was firstly founded in YM \cite{Hou:2018bwm}.  This relation states that a proper combination of Berends-Giele currents which are based on two connected tree graphs can be given by certain combination of products of two lower-point subcurrents corresponding to the two graphs. Although, the on-shell version of graph-based BCJ relation has been proven \cite{Hou:2018bwm} and has shown to be expanded in terms of usual BCJ relations \cite{Bern:2008qj,BjerrumBohr:2009rd,Stieberger:2009hq,Chen:2011jxa}, the off-shell relations \cite{Wu:2021bcy}, which was verified by several examples, have not been proven yet.   In this note, we prove the graph-based BCJ relation for the Berends-Giele currents in BS. The corresponding relation in YM theory is further induced by dressing the Berends-Giele currents in BS with proper numerators.

This note is organized as follows. In \secref{sec:Review_On_Basics}, we review the definition and properties of Berends-Giele currents in BS as well as  the off-shell extended graph-based BCJ relation. We further demonstrate simple examples in \secref{sec:examples}, which correspond to simple chains and star graphs. In \secref{sec:Proof_of_off-shell_BCJ_relation}, a general proof of the relation that is based on arbitrary tree graphs is provided. Further discussions including the on-shell limit, an antisymmetric form of the relation and the extensions to YM theory are given in \secref{sec:FurtherDiscussions}. We make a summary in \secref{sec:Conclusions}.

\section{Berends-Giele currents in BS theory and graph-based BCJ relation}\label{sec:Review_On_Basics}

In this section, we briefly review the definition and useful properties of Berends-Giele current in BS theory. The graph-based BCJ relation for BS currents is also introduced.

\subsection{Berends-Giele currents in BS theory}

In BS theory, the Berends-Giele current $\phi(1,2,\dots,n-1|\pmb{\sigma})$, where $\pmb{\sigma}$ is a permutation of $n-1$ elements (which are not necessarily the elements $1$, ..., $n-1$), can be recursively defined \cite{Mafra:2016ltu} by
\bea
\phi(1,2,\dots,n-1|\pmb{\sigma})
&\equiv&\frac{1}{s_{1\dots n-1}}\Sl_{1\leq i\leq n-2}\Bigl[\phi\left(1,2,\dots,i|\sigma_1,\dots,\sigma_i\right)\phi\left(i+1,\dots,n-1|\sigma_{i+1},\dots,\sigma_{n-1}\right) \nn
&&~-\phi\left(1,2,\dots,i|\sigma_{n-i},\dots,\sigma_{n-1}\right) \phi\left(i+1,\dots,n-1|\sigma_{1},\dots,\sigma_{n-i-1}\right) \Bigr]. \Label{Eq:BsBGcurrent}
\eea
In the above expression, $s_{12\dots n-1}\equiv(k_1+k_2+...+k_{n-1})^2$, where  $k_i$ is the momentum of the particle $i$ which satisfies the on-shell condition $k_i^2=0$. This recursive definition starts with the one-point current $\phi(l|l')$ that is defined by
\bea
\phi(l|l')=\Biggl\{
             \begin{array}{cc}
               1 & (l=l') \\
               0 & (l\neq l') \\
             \end{array}.
\eea
A consequent result of the definition is $\phi(a_1,\dots,a_i|b_1,\dots,b_i)=0$ if $\{a_1,\dots,a_i\}\setminus\{b_1,\dots,b_i\}\neq\emptyset$. The $n$-point on-shell partial amplitude $\mathcal{A}(1,2,\dots,n|\pmb{\sigma})$ is obtained by taking the following limit
\bea
\mathcal{A}(1,2,\dots,n|\pmb{\sigma})=s_{1\dots n-1}\phi(1,2,\dots,n-1|\pmb{\sigma})\big|_{k_n^2=s_{1\dots n-1}\rightarrow0}. \Label{Eq:BG-Amp}
\eea

Berends-Giele currents (\ref{Eq:BsBGcurrent}) in BS theory satisfy Kleiss-Kuijf (KK) relation \cite{Kleiss:1988ne}
  \bea
  &&\phi(1,2,\dots, n-1|\pmb{\beta},1,\pmb{\alpha}) =(-)^{|\pmb{\beta}|}\phi(1,2,\dots, n-1|1,\pmb{\alpha}\shuffle\pmb{\beta}^{T}), \Label{Eq:BsProperty1}
  \eea
  where $\pmb{\alpha}$, $\pmb{\beta}$ are two ordered sets, and $\pmb{\beta},1,\pmb{\alpha}$ is a permutation of $n-1$ elements. The notation $|\pmb{\beta}|$ stands for the number of elements in the ordered set $\pmb{\beta}$, and $\pmb{\beta}^T$  denotes the inverse permutation of $\pmb{\beta}$. For two ordered sets $\pmb{A}$ and $\pmb{B}$, $\pmb{A}\shuffle\pmb{B}$ denotes the set of all \emph{shuffle permutations}, which are obtained by merging $\pmb{A}$ and $\pmb{B}$ together with keeping the relative order of elements in each ordered set.  For example,   $\{2,4\}\shuffle\{3,5\}$ involves the following permutations
  \bea
  \{2,4,3,5\},\quad \{2,3,4,5\},\quad \{3,2,4,5\}, \quad\{2,3,4,5\},\quad\{3,2,5,4\},\quad\{3,5,2,4\}.
  \eea
 When the ordered set $\pmb{\alpha}$ in \eqref{Eq:BsProperty1} is empty, KK relation turns into the reflection relation
  \bea
  &&\phi(1,2,\dots, n-1|\sigma_1,\sigma_2,\dots,\sigma_{n-1})=(-)^{n}\phi\left(1,2,\dots, n-1|\sigma_{n-1},\sigma_{n-2},\dots,\sigma_{1}\right). \Label{Eq:BsProperty2}
  \eea
For given two ordered sets $\pmb{\alpha}\in \mathcal{P}(\{i_1,..,i_l\})$ and $\pmb{\beta}\in \mathcal{P}(\{1,2,...,n-1\}\setminus\{i_1,..,i_l\})$  where $i_1,..,i_l\in \{1,2,...,n-1\}$  and $\mathcal{P}(\{i_1,..,i_l\})$ denotes the set of all permutations of $i_1,..,i_l$, the Berends-Giele currents satisfy the following identity
  \bea
  &&\phi(1,2,\dots, n-1|\pmb{\alpha}\shuffle\pmb{\beta})=0, \Label{Eq:BsProperty3}
  \eea
  which is essentially equivalent to the KK relation  (\ref{Eq:BsProperty1}).

\subsection{Graph-based BCJ relation for BS currents}

\begin{figure}
  \centering
  \includegraphics[width=0.3\textwidth]{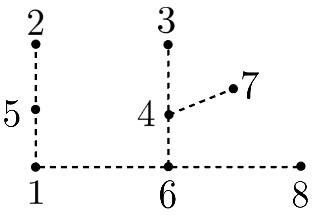}
  \caption{A connected tree graph $\mathcal{T}$ with eight nodes. When the nodes $1$ and $2$ are chosen as the leftmost elements respectively, the corresponding permutations are $\mathcal{T}\bigr|_{1}=\bigl\{1,\{5,2\}\shuffle\bigl\{6,\{4,3\shuffle7\}\shuffle8\bigr\}\bigr\}$ and $\mathcal{T}\bigr|_{2}=\bigl\{2,5,1,6,8\shuffle\{4,3\shuffle7\}\bigr\}$. }
  \label{Fig:example_graph}
\end{figure}
%
%
As pointed in \cite{Wu:2021bcy}, Berends-Giele currents  (\ref{Eq:BsBGcurrent})  satisfy so-called \emph{graph-based BCJ relation} whose on-shell version was first proposed for YM amplitudes \cite{Hou:2018bwm}. To show this relation,
 we introduce permutations $\pmb{\sigma}$ for a given connected tree graph $\mathcal{T}$ and a given node $a\in \mathcal{T}$ as follows: (i). the node  $a$ is considered as the leftmost element in $\pmb{\sigma}$, (ii). for any two adjacent elements $x$ and $y$, we have ${\sigma}^{-1}(x)< \sigma^{-1}(y)$ \footnote{The notation ${\sigma}^{-1}(x)$ denotes the position of the element $x$ in the permutation $\pmb{\sigma}$. In some places, we used $x \prec y$ to stand for ${\sigma}^{-1}(x)< \sigma^{-1}(y)$  for simplicity. }  if $x$ is nearer to $a$ than $y$, (iii). for subtrees attached to a same node, we shuffle the permutations corresponding to these subtrees together. The collection of all such permutations is denoted by $\mathcal{T}|_a$. For example, the set of permutations established by  \figref{Fig:example_graph} when the node $1$ is considered as the leftmost element  are given by
 \bea
\mathcal{T}|_1=\bigl\{1,\bigl\{\{5,2\}\shuffle \{6,\{4,\{3\}\shuffle\{7\}\}\shuffle\{8\}\}\bigr\}\bigr\}.
 \eea
Having the above definition, we define the following combination of BS currents which are characterized by two connected  tree graphs $\mathcal{T}_{\pmb {A}}$ and $\mathcal{T}_{\pmb {B}}$:
\bea
B^{(x)}(\pmb{\sigma}\,|\,\mathcal{T}_{\pmb {A}}|_a,\mathcal{T}_{\pmb{B}}|_b)
&\equiv&(-)^{|ax|}\Sl_{c\in \mathcal{T}_{\pmb{B}}}\Sl_{\pmb{\alpha}\in \mathcal{T}_{\pmb{A}}|_a}\Sl_{\pmb{\beta}\in \mathcal{T}_{\pmb{B}}|_b} \biggl[\,\Sl_{\pmb{\gamma}\in \pmb{\alpha}\shuffle\pmb{\beta}|_{c\prec a}}s_{ac}\,\phi(\,\pmb{\sigma}\,|\,\pmb{\gamma}\,)\,\biggr]. \Label{Eq:DefofBr}
\eea
where $s_{ac}=(k_{a}+k_{c})^2$.
On the LHS of the above expression, $x$ is an arbitrarily chosen node in $\mathcal{T}_{\pmb {A}}$, while $\pmb{\sigma}$ is an arbitrary permutation of all elements corresponding to nodes. On the RHS, the first summation is taken over all choices of $c$ in $\mathcal{T}_{\pmb{B}}$. The second and the third summations are taken over permutations established by $\mathcal{T}_{\pmb {A}}$ and $\mathcal{T}_{\pmb{B}}$ when $a$ and $b$ are considered as the leftmost elements, respectively. We further sum over all the shuffle permutations such that $c\prec a$ and associate a factor $s_{ac} $ to a given choice of $c$.
%
The $|ax|$ denotes the distance between $a$ and $x$ (the number of edges between $a$ and $x$).
According to this definition, the sign $(-)^{|ax|}$ can be determined as follows: (i). if  $a=x$, $(-)^{|ax|}=1$, (ii) for two adjacent nodes $y$ and $z$, $(-)^{|yx|}=-(-)^{|zx|}$. With the combination (\ref{Eq:DefofBr}), graph-based BCJ relation for BS currents is introduced as
\bea
\mathbb{B}^{(x)}(\pmb{\sigma}\,|\,\mathcal{T}_{\pmb {A}},\mathcal{T}_{\pmb{B}}|_b)
&\equiv&\Sl_{a\in \mathcal{T}_A}B^{(x)}\big(\pmb{\sigma}|\mathcal{T}_{\pmb {A}}|_a,\mathcal{T}_{\pmb{B}}|_b\bigr) \nn
&=&\Sl_{\pmb{\alpha}\in \mathcal{T}_{\pmb{A}}|_x}\Sl_{\pmb{\beta}\in \mathcal{T}_{\pmb{B}}|_b} \Big[\phi(\pmb{\sigma}_{1,i}|\pmb{\beta})\phi(\pmb{\sigma}_{i+1,l}|\pmb{\alpha}) -\phi(\pmb{\sigma}_{1,{l-i}}|\pmb{\alpha})\phi(\pmb{\sigma}_{l-i+1, l}|\pmb{\beta})\Big], \Label{Eq:OffBCJRelation}
\eea
%
where $\pmb{\sigma}_{i,j}$ ($i<j$) denotes $\{\sigma_i,\sigma_{i+1},...,\sigma_j\}$ for short, the  number of nodes in the two graphs $\mathcal{T}_{\pmb{A}}$ and $\mathcal{T}_{\pmb{B}}$ are supposed to be $i$ and $l-i$, respectively \footnote{ In \cite{Wu:2021bcy}, we have checked the relation up to six-point scattering.}.

\section{Examples of graph-based BCJ relation}\label{sec:examples}

In this section, we study explicit examples of the off-shell graph-based BCJ relation (\ref{Eq:OffBCJRelation}), which are established by graphs with simple structures.



\subsection*{Example-1}

The simplest example is the case in which each of $\mathcal{T}_{\pmb{A}}$ and $\mathcal{T}_{\pmb{B}}$ contains only one node, say $1$ and $2$, respectively. When the Berends-Giele recursion (\ref{Eq:BsBGcurrent}) for the BS current $\phi(\sigma_1,\sigma_2|2,1)$ is substituted into the LHS of (\ref{Eq:OffBCJRelation}), we get
\bea
\mathbb{B}^{(1)}(\sigma_1,\sigma_2\,|\,\mathcal{T}_{\pmb {A}}|_1,\mathcal{T}_{\pmb{B}}|_2)
&=&s_{12}\,\phi(\sigma_1,\sigma_2|2,1)\nn
&=&\bigl[\phi(\sigma_1|2)\phi(\sigma_2|1)-\phi(\sigma_1|1)\phi(\sigma_2|2)\bigr],\Label{Eq:Example1}
\eea
which reproduces the RHS of the graph-based BCJ relation (\ref{Eq:OffBCJRelation}) with  $\mathcal{T}_{\pmb{A}}=\{1\}$ and $\mathcal{T}_{\pmb{B}}=\{2\}$. 

%

\subsection*{Example-2}

\begin{figure}
  \centering
  \includegraphics[width=0.8\textwidth]{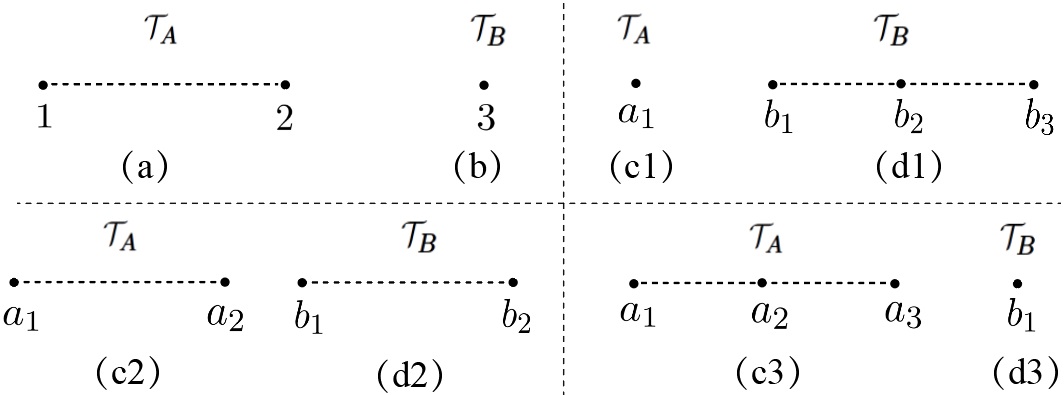}\\
  \caption{ In the relation for Berends-Giele currents with three external particles, if $\mathcal{T}_{\pmb{A}}$ involves two nodes and $\mathcal{T}_{\pmb{B}}$ has a single node, the possible structures of the graphs can only be (a) and (b). Permutations established by these graphs are $\mathcal{T}_{\pmb{A}}|_1=\{1,2\}$, $\mathcal{T}_{\pmb{A}}|_2=\{2,1\}$  and $\mathcal{T}_{\pmb{B}}|_3=\{3\}$. All possible graphs for the relation with four external particles are given by (c1), (d1), (c2), (d2) and (c3), (d3).}\label{Fig:ExampleThreePointBGcurrent}
\end{figure}

The second example is given by relations for three-point Berends-Giele currents. There are two possible cases corresponding to distinct number of nodes in the two graphs.

\paragraph{Case-1} If the graph  $\mathcal{T}_{\pmb{A}}$ involves two nodes, say $1$ and $2$, while the graph $\mathcal{T}_{\pmb{B}}$ contains a single node $3$, the structures of  $\mathcal{T}_{\pmb{A}}$ and   $\mathcal{T}_{\pmb{B}}$  are shown by \figref{Fig:ExampleThreePointBGcurrent} (a) and (b), respectively. The combinations of BS currents (\ref{Eq:DefofBr}) read:
\bea
B^{(1)}({\sigma}_1,\sigma_2,\sigma_3|\mathcal{T}_{\pmb{A}}\bigr|_{1},\mathcal{T}_{\pmb{B}}\bigr|_{3})
&=&\,\,\,s_{13}\,\phi({\sigma}_1,\sigma_2,\sigma_3|3,1,2), \Label{Eq:3ptExample5}\\
B^{(1)}({\sigma}_1,\sigma_2,\sigma_3|\mathcal{T}_{\pmb{A}}\bigr|_{2},\mathcal{T}_{\pmb{B}}\bigr|_{3})
&=&-s_{23}\,\phi({\sigma}_1,\sigma_2,\sigma_3|3,2,1). \Label{Eq:3ptExample6}
\eea
By using  Berends-Giele recursion (\ref{Eq:BsBGcurrent}), the current $\phi(1,2,3|3,1,2)$ will be decomposed into the following four terms
\bea
\phi(\sigma_1,\sigma_2,\sigma_3|3,1,2)
&=&\frac{1}{s_{123}}\Bigl[\phi(\sigma_1|3)\phi(\sigma_2,\sigma_3|1,2)+\phi(\sigma_1,\sigma_2|3,1)\phi(\sigma_3|2) \nn
&&~~~~~~~~~~~~-\phi(\sigma_3|3)\phi(\sigma_1,\sigma_2|1,2)-\phi(\sigma_2,\sigma_3|3,1)\phi(\sigma_1|2)\Bigr].
%
\eea
The expression of $\phi(\sigma_1,\sigma_2,\sigma_3|3,2,1)$ can be obtained via exchanging  $1$ and $2$ in the above equation.
When the Berends-Giele recursive expressions of $\phi(\sigma_1,\sigma_2,\sigma_3|3,1,2)$ and $\phi(\sigma_1,\sigma_2,\sigma_3|3,2,1)$ are substituted, the sum of \eqref{Eq:3ptExample5} and \eqref{Eq:3ptExample6} can be expressed by
\bea
&&\frac{s_{13}+s_{23}}{s_{123}}\Bigl[\phi(\sigma_1|3)\phi(\sigma_2,\sigma_3|1,2) -\phi(\sigma_1,\sigma_2|1,2)\phi(\sigma_3|3)\Bigr] \nn
&&~~~~~~~~~~+\frac{s_{13}}{s_{123}}\Bigl[\phi(\sigma_1,\sigma_2|3,1)\phi(\sigma_3|2) -\phi(\sigma_1|2)\phi(\sigma_2,\sigma_3|3,1)\Bigr] \nn
&&~~~~~~~~~~~~~-\frac{s_{23}}{s_{123}}\Bigl[\phi(\sigma_1,\sigma_2|3,2)\phi(\sigma_3|1) -\phi(\sigma_1|1)\phi(\sigma_2,\sigma_3|3,2) \Bigr].\Label{Eq:3ptExample7}
\eea
In the above expression, the coefficient on the first line is equal to $\frac{(s_{123}-s_{{12}})}{s_{123}}$. When the relation (\ref{Eq:Example1}) is applied to the two-point currents, the sum of the second and the third lines becomes
\bea
&&\frac{1}{s_{123}}\Bigl[\big(\phi(\sigma_1|3)\phi(\sigma_2|1)-\phi(\sigma_1|1)\phi(\sigma_2|3)\big)\phi(\sigma_3|2) -\phi(\sigma_1|2)\big(\phi(\sigma_2|3)\phi(\sigma_3|1)-\phi(\sigma_2|1)\phi(\sigma_3|3)\big)\Bigr] \nn
&-&\frac{1}{s_{123}}\Bigl[\big(\phi(\sigma_1|3)\phi(\sigma_2|2)-\phi(\sigma_1|2)\phi(\sigma_2|3)\big)\phi(\sigma_3|1) -\phi(\sigma_1|1)\big(\phi(\sigma_2|3)\phi(\sigma_3|2)-\phi(\sigma_2|2)\phi(\sigma_3|3)\big) \Bigr].
\eea
which further turns into $\frac{s_{12}}{s_{123}}\Bigl[\phi(\sigma_1|3)\phi(\sigma_2,\sigma_3|1,2) -\phi(\sigma_1,\sigma_2|1,2)\phi(\sigma_3|3)\Bigr]$. Therefore, the total contribution of \eqref{Eq:3ptExample5} and \eqref{Eq:3ptExample6} becomes
\bea
\mathbb{B}^{(1)}({\sigma}_1,\sigma_2,\sigma_3\,|\,\mathcal{T}_{\pmb {A}}|_1,\mathcal{T}_{\pmb{B}}|_3)&=&B^{(1)}({\sigma}_1,\sigma_2,\sigma_3|\mathcal{T}_{\pmb{A}}\bigr|_{1},\mathcal{T}_{\pmb{B}}\bigr|_{3})+B^{(1)}({\sigma}_1,\sigma_2,\sigma_3|\mathcal{T}_{\pmb{A}}\bigr|_{2},\mathcal{T}_{\pmb{B}}\bigr|_{3})\nn
&=&\biggl[\phi(\sigma_1|3)\phi(\sigma_2,\sigma_3|1,2)-\phi(\sigma_1,\sigma_2|1,2)\phi(\sigma_3|3) \biggr],
\eea
which is just the off-shell graph-based BCJ relation  (\ref{Eq:OffBCJRelation}) for the graphs $\mathcal{T}_{\pmb{A}}=\text{\figref{Fig:ExampleThreePointBGcurrent} (a)}$ and $\mathcal{T}_{\pmb{B}}=\text{\figref{Fig:ExampleThreePointBGcurrent} (b)}$.

\paragraph{Case-2} Now we exchange the roles between the two graphs in the above case, the LHS of (\ref{Eq:OffBCJRelation}) is then written as
\bea
\mathbb{B}^{(3)}({\sigma}_1,\sigma_2,\sigma_3\,|\,\mathcal{T}_{\pmb{B}}|_3,\mathcal{T}_{\pmb {A}}|_1)
&=&s_{13}\,\phi(\sigma_1,\sigma_2,\sigma_3|1,3,2) +(s_{13}+s_{23})\,\phi(\sigma_1,\sigma_2,\sigma_3|1,2,3).\Label{Eq:3ptExample8}
\eea
Expressing the BS currents on the RHS by Berends-Giele recursion,
we get
\bea
\mathbb{B}^{(3)}({\sigma}_1,\sigma_2,\sigma_3\,|\,\mathcal{T}_{\pmb{B}}|_3,\mathcal{T}_{\pmb {A}}|_1)=-\biggl[\phi(\sigma_1|3)\phi(\sigma_2,\sigma_3|1,2)-\phi(\sigma_1,\sigma_2|1,2)\phi(\sigma_3|3) \biggr],
\eea
which is the expected graph-based BCJ relation (\ref{Eq:OffBCJRelation}) with $\mathcal{T}_{\pmb {A}}=\text{\figref{Fig:ExampleThreePointBGcurrent} (b)}$ and $\mathcal{T}_{\pmb {B}}=\text{\figref{Fig:ExampleThreePointBGcurrent} (a)}$. From case-1 and case-2, we can see   the RHS is not changed upto a sign when we exchange the roles of the two graphs $\mathcal{T}_{\pmb {A}}$ and $\mathcal{T}_{\pmb {B}}$.
%
%
%

If the Berends-Giele currents in \eqref{Eq:OffBCJRelation} involve four external particles, three possibilities should be considered according to distinct structures of the pair of graphs $\mathcal{T}_{\pmb{A}}$ and $\mathcal{T}_{\pmb{B}}$, i.e., \figref{Fig:ExampleThreePointBGcurrent} (c1), (d1), (c2), (d2) and  (c3), (d3). In all these cases, graphs are simple chains (which are defined as graphs with no branch). We now skip the four-point cases and turn to an example of $5$ external particles, which involves a star structure rather than simple chain.

\subsection*{Example-3}

\begin{figure}
  \centering
  \includegraphics[width=0.8\textwidth]{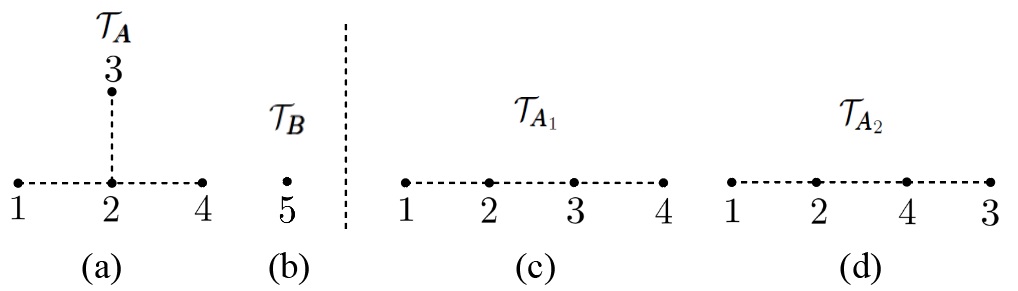}\\
  \caption{The graph (a) is a star graph while (b) is a single node. If the $\mathcal{T}_{\pmb{A}}$ and $\mathcal{T}_{\pmb{B}}$ in the graph-based BCJ relation are respectively chosen as (a) and (b),  the $\mathcal{T}_{\pmb{A}}$ can be decomposed into two simple chains (c) and (d), respectively. } \label{Fig:StarGraphExample}
\end{figure}

Now let us study the example in which  $\mathcal{T}_{\pmb{A}}$ corresponds to the star graph \figref{Fig:StarGraphExample} (a), while $\mathcal{T}_{\pmb{B}}$ refers to the single node $5$.  Although the relation (\ref{Eq:OffBCJRelation}) in this case can also be directly verified  via Berends-Giele recursion, we illustrate this example in another way, by expanding the combinations corresponding to the start graph in terms of those corresponding to simple chains.
Particularly, the LHS of (\ref{Eq:OffBCJRelation}) for $\mathcal{T}_{{A}}=\text{\figref{Fig:StarGraphExample} (a)}$  and  $\mathcal{T}_{{B}}=\text{\figref{Fig:StarGraphExample} (b)}$ can be expanded as
\bea
\mathbb{B}^{(1)}(\pmb{\sigma}\,|\,\mathcal{T}_{\pmb {A}},\mathcal{T}_{\pmb{B}}|_b)=\mathbb{B}^{(1)}(\pmb{\sigma}\,|\,\mathcal{T}_{\pmb {A}_1},\mathcal{T}_{\pmb{B}}|_b)+\mathbb{B}^{(1)}(\pmb{\sigma}\,|\,\mathcal{T}_{\pmb {A}_2},\mathcal{T}_{\pmb{B}}|_b),
\eea
where $\pmb{\sigma}$ stands for a permutation with the five external particles, while $\mathcal{T}_{\pmb {A}_1}$ and $\mathcal{T}_{\pmb {A}_2}$ are the two simple chains, as shown by \figref{Fig:StarGraphExample} (c), (d). This can be easily verified by writing down the permutations and signs corresponding to $\mathcal{T}_{\pmb {A}}$, $\mathcal{T}_{\pmb {A}_1}$ and $\mathcal{T}_{\pmb {A}_2}$ explicitly
\bea
\mathcal{T}_{\pmb {A}}|_1&\to&(+)\{1,2,\{3\}\shuffle\{4\}\}~~~~~~~\,\mathcal{T}_{\pmb {A}_1}|_1\to(+)\{1,2,3,4\}~~~~~~~~~~~\mathcal{T}_{\pmb {A}_2}|_1\to(+)\{1,2,4,3\}\nn
\mathcal{T}_{\pmb {A}}|_2&\to&(-)\{2,\{1\}\shuffle\{3\}\shuffle\{4\}\}~~\mathcal{T}_{\pmb {A}_1}|_2\to(-)\{2,\{1\}\shuffle\{3,4\}\}~~\mathcal{T}_{\pmb {A}_2}|_2\to(+)\{2,\{1\}\shuffle\{4,3\}\}\nn
\mathcal{T}_{\pmb {A}}|_3&\to&(+)\{3,2,\{1\}\shuffle\{4\}\}~~~~~~~\,\mathcal{T}_{\pmb {A}_1}|_3\to(+)\{3,\{2,1\}\shuffle\{4\}\}~~\,\mathcal{T}_{\pmb {A}_1}|_3\to(-)\{3,4,2,1\}\nn
\mathcal{T}_{\pmb {A}}|_4&\to&(+)\{4,2,\{1\}\shuffle\{3\}\}~~~~~~~\,\mathcal{T}_{\pmb {A}_1}|_4\to(-)\{4,3,2,1\}~~~~~~~~~~\,\,\mathcal{T}_{\pmb {A}_1}|_3\to(+)\{4,\{2,1\}\shuffle\{3\}\}.\Label{Eq:Example3Permutations}
\eea
Once $\mathbb{B}^{(1)}(\pmb{\sigma}\,|\,\mathcal{T}_{\pmb {A}_1},\mathcal{T}_{\pmb{B}}|_b)$ and $\mathbb{B}^{(1)}(\pmb{\sigma}\,|\,\mathcal{T}_{\pmb {A}_2},\mathcal{T}_{\pmb{B}}|_b)$, where $\mathcal{T}_{\pmb {A}_1}$ and $\mathcal{T}_{\pmb {A}_2}$ are the simple chains \figref{Fig:StarGraphExample} (c) and (d)  respectively, satisfy the relation (\ref{Eq:OffBCJRelation}), we have
\bea
\mathbb{B}^{(1)}(\pmb{\sigma}\,|\,\mathcal{T}_{\pmb {A}_1},\mathcal{T}_{\pmb{B}}|_b)&=&\Sl_{\pmb{\alpha}\in \mathcal{T}_{\pmb {A}_1}|_1}\Big[\phi(\pmb{\sigma}_{1}|5)\phi(\pmb{\sigma}_{2,5}|\pmb{\alpha}) -\phi(\pmb{\sigma}_{1,{4}}|\pmb{\alpha})\phi(\pmb{\sigma}_{5}|5)\Big]\nn
\mathbb{B}^{(1)}(\pmb{\sigma}\,|\,\mathcal{T}_{\pmb {A}_2},\mathcal{T}_{\pmb{B}}|_b)&=&\Sl_{\pmb{\alpha}\in \mathcal{T}_{\pmb {A}_2}|_1}\Big[\phi(\pmb{\sigma}_{1}|5)\phi(\pmb{\sigma}_{2,5}|\pmb{\alpha}) -\phi(\pmb{\sigma}_{1,{4}}|\pmb{\alpha})\phi(\pmb{\sigma}_{5}|5)\Big].
\eea
According to the first line of (\ref{Eq:Example3Permutations}), the sum of $\mathbb{B}^{(1)}(\pmb{\sigma}\,|\,\mathcal{T}_{\pmb {A}_1},\mathcal{T}_{\pmb{B}}|_b)$ and $\mathbb{B}^{(1)}(\pmb{\sigma}\,|\,\mathcal{T}_{\pmb {A}_2},\mathcal{T}_{\pmb{B}}|_b)$ precisely reproduces
\bea
\Sl_{\pmb{\alpha}\in \mathcal{T}_{\pmb {A}}|_1}\Big[\phi(\pmb{\sigma}_{1}|5)\phi(\pmb{\sigma}_{2,5}|\pmb{\alpha}) -\phi(\pmb{\sigma}_{1,{4}}|\pmb{\alpha})\phi(\pmb{\sigma}_{5}|5)\Big],
\eea
which is the RHS of (\ref{Eq:OffBCJRelation}) with $\mathcal{T}_{\pmb {A}}=\text{\figref{Fig:StarGraphExample} (a)}$ and $\mathcal{T}_{\pmb {B}}=\text{\figref{Fig:StarGraphExample} (b)}$.

\section{General proof of the graph-based BCJ relation}\label{sec:Proof_of_off-shell_BCJ_relation}

What we can learn from the previous section is that the relation based on graphs with branches (e.g. the star graph in \figref{Fig:StarGraphExample} (a)) can be related to relations with simple chains (e.g. the graphs \figref{Fig:StarGraphExample} (c), (d)). Inspired by this observation, we now prove the general relation  by (i) proving the relation (\ref{Eq:OffBCJRelation}) with two simple chains $\mathcal{T}_{\pmb{A}}$ and $\mathcal{T}_{\pmb{B}}$, via Berends-Giele recursion and then (ii) proving the relation (\ref{Eq:OffBCJRelation}) with arbitrary tree structures $\mathcal{T}_{\pmb{A}}$ and $\mathcal{T}_{\pmb{B}}$, via expanding the combinations on both sides in terms of those corresponding to simple chains.

\begin{figure}
  \centering
  \includegraphics[width=0.75\textwidth]{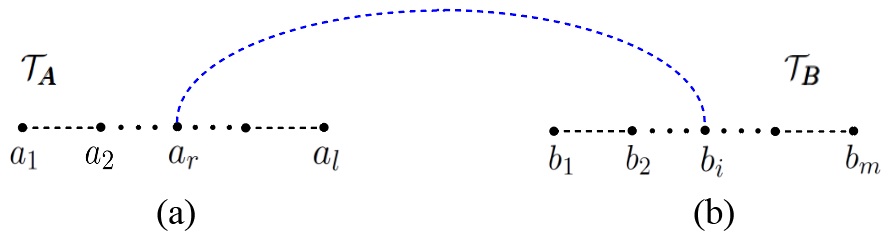}
  \caption{  In this graph,  $\mathcal{T}_{\pmb{A}}$ and $\mathcal{T}_{\pmb{B}}$ are both simple chains. The permutations for given choice of $a_r$ and $b_i$ in the graph-based BCJ relation can be conveniently determined by connecting $a_r$ and $b_i$ via a line, and the node $b_1$ is always regardered as the leftmost element.  }
  \label{Fig:SimpleChainOffShellBCJgraph}
\end{figure}
\subsection{Graph-based BCJ relation  with simple chains }\label{sec:AlphaAsPermutation}

When both $\mathcal{T}_{\pmb{A}}$ and $\mathcal{T}_{\pmb{B}}$ stand for simple chains, as shown by \figref{Fig:SimpleChainOffShellBCJgraph},  the LHS of \eqref{Eq:OffBCJRelation} are expressed by
\bea
\mathbb{B}^{(a_1)}\left(\pmb{\sigma}\,|\,\mathcal{T}_{\pmb {A}},\mathcal{T}_{\pmb{B}}|_{b_1}\right)&=&\Sl_{r=1}^{l}(-1)^{r-1}\Sl_{i=1}^{m}(2k_{a_r}\cdot k_{b_i})\,\Sl_{\pmb{\gamma}}\phi\left(\pmb{\sigma}|\pmb{\gamma}\right),\Label{Eq:Gen1}
\eea
where we have written $s_{a_r,b_i}$ by $2k_{a_r}\cdot k_{b_i}$ explicitly. The permutations $\pmb{\gamma}$ for given $i$, $r$ are defined by
\bea
\pmb{\gamma}\in\big\{b_1,...,b_i,\{a_{r},\{a_{r+1},...,a_l\}\shuffle\{a_{r-1},...,a_1\}\}\shuffle\{b_{i+1},...,b_m\}\big\}.\Label{Eq:gamma}
\eea
Noting that one can exchange the summations over $i$ and $\pmb{\gamma}$ in \eqref{Eq:Gen1} by collecting the coefficients corresponding to a given $\pmb{\gamma}$ and then summing over all possible permutations. As a result, $\mathbb{B}^{(a_1)}\left(\pmb{\sigma}\,|\,\mathcal{T}_{\pmb {A}},\mathcal{T}_{\pmb{B}}|_{b_1}\right)$ is reexpressed by
\bea
\mathbb{B}^{(a_1)}\left(\pmb{\sigma}\,|\,\mathcal{T}_{\pmb {A}},\mathcal{T}_{\pmb{B}}|_{b_1}\right)&=&\Sl_{r=1}^{l}(-1)^{r-1}\Sl_{\pmb{\gamma}}2k_{a_r}\cdot\Big(\Sl_{b_i\prec a_r} k_{b_i}\Big)\,\phi\left(\pmb{\sigma}|\pmb{\gamma}\right),\Label{Eq:Gen2}
\eea
where we have summed over all possible $\pmb{\gamma}$ satisfying
\bea
\pmb{\gamma}\in\big\{b_1,\{b_2,...,b_m\}\shuffle\{a_r,\{a_{r+1},...,a_l\}\shuffle\{a_{r-1},...,a_1\}\}\big\}.\Label{Eq:gamma2}
\eea
Expressing each current in \eqref{Eq:Gen2} by Berends-Giele recursion (\ref{Eq:BsBGcurrent}), we rewrite \eqref{Eq:Gen2} as
\bea
&&~~~\mathbb{B}^{(a_1)}\left(\pmb{\sigma}\,|\,\mathcal{T}_{\pmb {A}},\mathcal{T}_{\pmb{B}}|_{b_1}\right)\Label{Eq:Gen3}\nn
&=&\Sl_{r=1}^{l}(-1)^{r-1}{1\over s_{1...n-1}}\biggl\{\Sl_{\pmb{\gamma}}2k_{a_r}\cdot\Big(\Sl_{b_i\prec a_r} k_{b_i}\Big)\Sl_{j=1}^{m}\Big[\,\phi\left(\pmb{\sigma}_{1,j}|\pmb{\gamma}_{1,j}\right)\phi\left(\pmb{\sigma}_{j+1,n-1}|\pmb{\gamma}_{j+1,n-1}\right)\nn
&&~~~~~~~~~~~~~~~~~~~~~~~~~~~~~~~~~~~~~~~~~~~~~~~~~~~~~~~\,\,\,\,-\phi\left(\pmb{\sigma}_{1,j}|\pmb{\gamma}_{n-j,n-1}\right)\phi\left(\pmb{\sigma}_{j+1,n-1}|\pmb{\gamma}_{1,(n-1)-j}\right)\Big]\biggr\},
\eea
where $n-1=l+m$ is the total number of on-shell particles. One can collect those subcurrents with the same particle contents but distinct permutations together. The contributions of the first term in the above expression can be collected as follows, according to whether $a_r\in\pmb{\gamma}_{1,j}$ or $a_r\in \pmb{\gamma}_{j+1,n-1}$.

\begin{figure}
  \centering
  \includegraphics[width=0.9\textwidth]{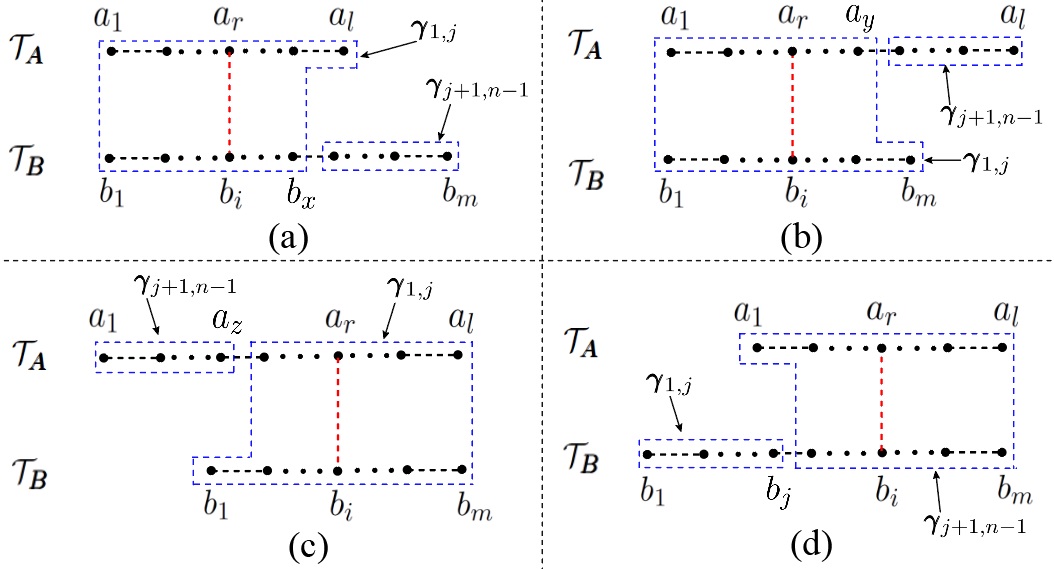}
  \caption{ Four types of graphs contributing to (\ref{Eq:Gen4}). In each graph, the boxed subgraph containing the element $b_1$ contributes to permutations $\pmb{\gamma}_{1,j}$ in (\ref{Eq:Gen4}) while the other boxed subgraph contributes to $\pmb{\gamma}_{j+1,n-1}$. }
  \label{Fig:Decomposition}
\end{figure}

\noindent {\bf(i).} When $a_r\in \pmb{\gamma}_{1,j}$, the permutations $\pmb{\gamma}_{1,j}$ and $\pmb{\gamma}_{j+1,n-1}$ in general have the following forms
\bea
\pmb{\gamma}_{1,j}&\in&\bigl\{b_1,\{b_2,...,b_x\}\shuffle\{a_r,\{a_{r+1},a_{r+2},...,a_y\}\shuffle\{a_{r-1},a_{r-2},...,a_z\}\}\bigr\},\Label{Eq:gamma3}\\
\pmb{\gamma}_{j+1,n-1}&\in&\{b_{x+1},b_{x+2},...,b_m\}\shuffle\{a_{y+1},a_{y+2},...,a_l\}\shuffle\{a_{z},a_{z-1},...,a_1\},\Label{Eq:gamma4}
\eea
where $x$, $y$ and $z$ satisfy the following constraint
\bea
x+(y-r+1)+(r-z)=j.
\eea
The situations in which one or two of the ordered sets $\{b_{x+1},b_{x+2},...,b_m\}$, $\{a_{y+1},a_{y+2},...,a_l\}$ and $\{a_{z},a_{z-1},...,a_1\}$ are empty are considered as boundary cases. The total contribution of the first term in \eqref{Eq:Gen3} is then collected as
\bea
{1\over s_{1...n-1}}\Sl_{j=1}^{m}\left\{\Sl_{r=1}^{l}(-1)^{r-1}\bigg[\,\Sl_{\pmb{\gamma}_{1,j}}2k_{a_r}\cdot\Big(\Sl_{b_i\prec a_r} k_{b_i}\Big)\phi\left(\pmb{\sigma}_{1,j}|\pmb{\gamma}_{1,j}\right)\bigg]\right\}\bigg[\Sl_{\pmb{\gamma}_{j+1,n-1}}\phi\left(\pmb{\sigma}_{j+1,n-1}|\pmb{\gamma}_{j+1,n-1}\right)\bigg].\Label{Eq:Gen4}
\eea
If more than two of the ordered sets $\{b_{x+1},b_{x+2},...,b_m\}$, $\{a_{y+1},a_{y+2}...,a_l\}$ and $\{a_{z},a_{z-1}...,a_1\}$ are nonempty, the factor in the last square brackets has to vanish, due to the generalized KK relation (\ref{Eq:BsProperty3}). Therefore, only those boundary cases when one of the ordered sets in \eqref{Eq:gamma4} is nonempty survive, as shown by \figref{Fig:Decomposition}.
\begin{itemize}
\item If $\pmb{\gamma}_{j+1,n-1}=\{b_{x+1},b_{x+2},...,b_m\}$, the second factor in \eqref{Eq:Gen4} reads
\bea
\phi\left(\pmb{\sigma}_{l+x+1,n-1}|b_{x+1},b_{x+2},...,b_m\right),
\eea
while the first factor is just given by \eqref{Eq:Gen2}  via removing the nodes $b_{x+1},b_{x+2},...,b_m$ from the chain  $\mathcal{T}_{\pmb{B}}$ (see \figref{Fig:Decomposition} (a)). According to the graph-based BCJ relation for lower-point currents, this factor is written as
\bea
\Big[\phi(\pmb{\sigma}_{1,x}|b_{1},b_{2},...,b_x)\phi(\pmb{\sigma}_{x+1,x+l}|a_1,a_2,...,a_l)-\phi(\pmb{\sigma}_{1,l}|a_1,a_2,...,a_l)\phi(\pmb{\sigma}_{l+1,l+x}|b_{1},b_{2},...,b_x)\Big].
\eea
Collecting the two factors together and summing over all possible $x$,  we rewrite the expression (\ref{Eq:Gen4}) as
\bea
T_1&=&{1\over s_{1...n-1}}\Sl_{1\leq x< m}\Big[\phi(\pmb{\sigma}_{1,x}|b_{1},b_{2},...,b_x)\phi(\pmb{\sigma}_{x+1,x+l}|a_1,a_2,...,a_l)\\
&&~~~~~~~~~~~~~~\,\,-\phi(\pmb{\sigma}_{1,l}|a_1,a_2,...,a_l)\phi(\pmb{\sigma}_{l+1,x+l}|b_{1},b_{2},...,b_x)\Big]\phi\left(\pmb{\sigma}_{l+x+1,n-1}|b_{x+1},b_{x+2},...,b_m\right).\nonumber
\eea
\item If $\pmb{\gamma}_{j+1,n-1}=\{a_{y+1},a_{y+2}...,a_l\}$, the corresponding graph is given by \figref{Fig:Decomposition} (b).  The last factor in (\ref{Eq:Gen4}) becomes
\bea
\phi\left(\pmb{\sigma}_{m+y+1,n-1}|a_{y+1},a_{y+2}...,a_l\right),
\eea
while the factor inside the braces turns into
\bea
&&\Big[\phi(\pmb{\sigma}_{1,m}|b_{1},b_{2},...,b_m)\phi(\pmb{\sigma}_{m+1,m+y}|a_1,a_2,...,a_y)\nn
&&~~~~~~~~~~~~~~~~~~~~~~~-\phi(\pmb{\sigma}_{1,y}|a_1,a_2,...,a_y)\phi(\pmb{\sigma}_{y+1,m+y}|b_{1},b_{2},...,b_m)\Big].
\eea
The total contribution of this case is  therefore obtained by summing over $y$ as follows
\bea
T_2&=&{1\over s_{1...n-1}}\Sl_{1\leq y< l}\Big[\phi(\pmb{\sigma}_{1,m}|b_{1},b_{2},...,b_m)\phi(\pmb{\sigma}_{m+1,m+y}|a_1,a_2,...,a_y)\\
&&~~~~~~~~~\,\,-\phi(\pmb{\sigma}_{1,y}|a_1,a_2,...,a_y)\phi(\pmb{\sigma}_{y+1,m+y}|b_{1},b_{2},...,b_m)\Big]\phi\left(\pmb{\sigma}_{m+y+1,n-1}|a_{y+1},a_{y+2},...,a_l\right).\nonumber
\eea

\item If $\pmb{\gamma}_{j+1,n-1}=\{a_{z},a_{z-1},...,a_1\}$, the corresponding graph is given by \figref{Fig:Decomposition} (c) and the last factor in (\ref{Eq:Gen4}) becomes
\bea
\phi\left(\pmb{\sigma}_{n-z,n-1}|a_{z},a_{z-1},...,a_1\right),
\eea
while the factor inside the braces becomes
\bea
&&\Big[\phi(\pmb{\sigma}_{1,m}|b_{1},b_{2},...,b_m)\phi(\pmb{\sigma}_{m+1,(n-1)-z}|a_{z+1},a_{z+2},...,a_l)\nn
&&~~~~~~~~~~~~~~~~~\,-\phi(\pmb{\sigma}_{1,l}|a_{z+1},a_{z+2},...,a_l)\phi(\pmb{\sigma}_{l+1,(n-1)-z}|b_{1},b_{2},...,b_m)\Big].
\eea
The total contribution of this part is then collected by summing over $z$:
\bea
T_3&=&{1\over s_{1...n-1}}\Sl_{1\leq z<l}\Big[\phi(\pmb{\sigma}_{1,m}|b_{1},b_{2},...,b_m)\phi(\pmb{\sigma}_{m+1,(n-1)-z}|a_{z+1},a_{z+2},...,a_l)\\
&&~~~~~~~~~~~\,-\phi(\pmb{\sigma}_{1,l-z}|a_{z+1},a_{z+2},...,a_l)\phi(\pmb{\sigma}_{l-z+1,(n-1)-z}|b_{1},b_{2},...,b_m)\Big]\phi\left(\pmb{\sigma}_{n-z,n-1}|a_{1},a_{2},...,a_z\right).\nonumber
\eea
\end{itemize}

\noindent{\bf(ii).} When $a_r\in\pmb{\gamma}_{j,n-1}$, a typical graph is given by \figref{Fig:Decomposition} (d). The permutations $\pmb{\gamma}_{1,j}$ and $\pmb{\gamma}_{j+1,n-1}$ satisfy
\bea
\pmb{\gamma}_{1,j}&=&\{b_1,b_2,...,b_j\},\Label{Eq:gamma5}\\
\pmb{\gamma}_{j+1,n-1}&\in&\{b_{j+1},b_{j+2},...,b_m\}\shuffle\{a_{r},\{a_{r+1},...,a_l\}\shuffle\{a_{r-1},...,a_1\}\}.\Label{Eq:gamma6}
\eea
The total contribution of the first term in \eqref{Eq:Gen3}  in this case can be expressed as
\bea
&&{1\over s_{1...n-1}}\Sl_{j=1}^{m}\phi\left(\pmb{\sigma}_{1,j}|b_1,b_2,...,b_j\right)\Biggl[\,\Sl_{r=1}^{l}(-1)^{r-1}\Sl_{\pmb{\gamma}_{j+1,n-1}}\Big(\Sl_{\substack{b_i\in\{b_{j+1},...,b_m\}\\ b_i\prec a_r}}s_{a_r,b_i} \Big)\phi\left(\pmb{\sigma}_{j+1,n-1}|\pmb{\gamma}_{j+1,n-1}\right)\nn
&&~~~~~~~~~~~~~~~~~~~~~~~~~~~~~~~~~~~~+\Sl_{r=1}^{l}(-1)^{r-1}\Big(\Sl_{b_i\in\{b_1,...,b_j\}} s_{a_r,b_i} \Big)\Sl_{\pmb{\gamma}_{j+1,n-1}}\phi\left(\pmb{\sigma}_{j+1,n-1}|\pmb{\gamma}_{j+1,n-1}\right)
\Biggr].\Label{Eq:Gen5}
\eea
When the graph-based BCJ relation is applied, the total contribution of the first term in  \eqref{Eq:Gen5} reads
\bea
T_4&=&{1\over s_{1...n-1}}\Sl_{j=1}^{m}\phi\left(\pmb{\sigma}_{1,j}|b_1,b_2,...,b_j\right)\Big[\phi\left(\pmb{\sigma}_{j+1,m}|b_{j+1},...,b_m\right)\phi(\pmb{\sigma}_{m+1,m+l}|a_1,...,a_l)\nn
&&~~~~~~~~~~~~~~~~~~~~~~~~~~~~~~~~~~~~~~~~~-\phi\left(\pmb{\sigma}_{j+1,j+l}|a_1,...,a_l\right)\phi(\pmb{\sigma}_{j+l+1,m+l}|b_{j+1},...,b_m)\Big].
\eea
The second term in the square brackets of \eqref{Eq:Gen5} has to vanish due to the generalized KK relation (\ref{Eq:BsProperty3}) if $j<m$. In the boundary case $j=m$, $\pmb{\gamma}_{j+1,n-1}=\{a_{r},\{a_{r+1},...,a_l\}\shuffle\{a_{r-1},...,a_1\}\}$, KK relation allows one to rewrite the summation over $\pmb{\gamma}_{j+1,n-1}$ as
\bea
\Sl_{\pmb{\gamma}_{j+1,n-1}}\phi\left(\pmb{\sigma}_{j+1,n-1}|\pmb{\gamma}_{j+1,n-1}\right)=(-1)^{r-1}\phi\left(\pmb{\sigma}_{j+1,n-1}|a_1,...,a_l\right).
\eea
Hence the total contribution of the second term in \eqref{Eq:Gen5} turns into
\bea
T_5&=&{1\over s_{1...n-1}}\Sl_{r=1}^{l}\Sl_{i=1}^m(2k_{a_r}\cdot k_{b_i})\phi\left(\pmb{\sigma}_{1,m}|b_1,...,b_m\right)\phi\left(\pmb{\sigma}_{j+1,n-1}|a_1,...,a_l\right)\nn
&=&{1\over s_{1...n-1}}\left(s_{1...n-1}-s_{a_1...a_l}-s_{b_1...b_m}\right)\phi\left(\pmb{\sigma}_{1,m}|b_1,...,b_m\right)\phi\left(\pmb{\sigma}_{j+1,n-1}|a_1,...,a_l\right).
\eea

Exchanging the roles of $\pmb{\sigma}_{1,j}$ and $\pmb{\sigma}_{j+1,n-1}$ and repeating the above discussions, we get the corresponding contributions of the second term in \eqref{Eq:Gen3}, say, $T'_1$, $T'_2$, $T'_3$, $T'_4$ and $T'_5$:
\bea
T'_1&=&{1\over s_{1...n-1}}\Sl_{1\leq x< m}\phi(\pmb{\sigma}_{1,x}|b_{m-x+1},b_{m-x+2},...,b_m)\Big[\phi(\pmb{\sigma}_{x+1,m}|b_1,b_2,...,b_{m-x})\phi(\pmb{\sigma}_{m+1,n-1}|a_1,a_2,...,a_l)\nn
&&~~~~~~~~~~~~~~~~~~~~-\phi(\pmb{\sigma}_{x+1,x+l}|a_1,a_2,...,a_{l})\phi(\pmb{\sigma}_{x+l+1,n-1}|b_1,b_2,...,b_{m-x})\Big],\\
T_2'&=&{1\over s_{1...n-1}}\Sl_{1\leq y< l}\phi(\pmb{\sigma}_{1,l-y}|a_{y+1},a_{y+2},...,a_l)\Big[\phi(\pmb{\sigma}_{l-y+1,(n-1)-y}|b_{1},b_{2},...,b_m)\phi(\pmb{\sigma}_{n-y,n-1}|a_1,a_2,...,a_y)\nn
&&~~~~~~~~~~~~~~~~~~~~~~~~~~~~~-\phi(\pmb{\sigma}_{l-y+1,l}|a_1,a_2,...,a_y)\phi(\pmb{\sigma}_{l+1,n-1}|b_{1},b_{2},...,b_m)\Big],\\
T_3'&=&{1\over s_{1...n-1}}\Sl_{1\leq z<l}\phi(\pmb{\sigma}_{1,z}|a_{1},a_{2},...,a_z)\Big[\phi(\pmb{\sigma}_{z+1,z+m}|b_{1},b_{2},...,b_m)\phi(\pmb{\sigma}_{z+m+1,n-1}|a_{z+1},a_{z+2},...,a_l)\nn
&&~~~~~~~~~~~~~~~~~~~~~~~~~~~~~-\phi(\pmb{\sigma}_{z+1,l}|a_{z+1},...,a_l)\phi(\pmb{\sigma}_{l+1,n-1}|b_{1},b_{2},...,b_m)\Big],\\
T_4'&=&{1\over s_{1...n-1}}\Sl_{j=1}^{m}\Big[\phi\left(\pmb{\sigma}_{1,m-j}|b_{j+1},...,b_m\right)\phi(\pmb{\sigma}_{m-j+1,(n-1)-j}|a_1,...,a_l)\nn
&&~~~~~~~~~~~~~~~~~~~~~~~~~~-\phi\left(\pmb{\sigma}_{1,l}|a_1,...,a_l\right)\phi(\pmb{\sigma}_{l+1,(n-1)-j}|b_{j+1},...,b_m)\Big]\phi\left(\pmb{\sigma}_{n-j,n-1}|b_1,...,b_j\right)\\
T_5'&=&{1\over s_{1...n-1}}\left(s_{1...n-1}-s_{a_1...a_l}-s_{b_1...b_m}\right)\phi\left(\pmb{\sigma}_{1,l}|a_1,...,a_l\right)\phi\left(\pmb{\sigma}_{l+1,n-1}|b_1,...,b_m\right).
\eea
The $\mathbb{B}^{(a_1)}\left(\pmb{\sigma}\,|\,\mathcal{T}_{\pmb {A}},\mathcal{T}_{\pmb{B}}|_{b_1}\right)$ in \eqref{Eq:Gen3} is finally written as
\bea
\mathbb{B}^{(a_1)}\left(\pmb{\sigma}\,|\,\mathcal{T}_{\pmb {A}},\mathcal{T}_{\pmb{B}}|_{b_1}\right)=\Sl_{i=1}^5\left[T_i-T_i'\right]. \Label{BTerm}
\eea

To sum up, the five terms are collected as
\bea
I_1+I_2+T_5-T_5'\Label{Eq:Summation}
\eea
where the $I_1$ and $I_2$ in the above expression are given by
\bea
I_1&\equiv&(T_1-T_1')+(T_4-T_4') \nn
&=&\frac{s_{b_1...b_m}}{s_{1...n-1}}\left[
\phi\left(\pmb{\sigma}_{1,m}|b_1,...,b_m\right)\phi\left(\pmb{\sigma}_{m+1,n-1}|a_1,...,a_l\right)
-\phi\left(\pmb{\sigma}_{1,l}|a_1,...,a_l\right)\phi\left(\pmb{\sigma}_{l+1,n-1}|b_1,...,b_m\right) \right],\Label{Eq:I1} \\
I_2&\equiv&(T_2-T_2')+(T_3-T_3') \nn
&=&\frac{s_{a_1...a_m}}{s_{1...n-1}}\left[
\phi\left(\pmb{\sigma}_{1,m}|b_1,...,b_m\right)\phi\left(\pmb{\sigma}_{m+1,n-1}|a_1,...,a_l\right)
-\phi\left(\pmb{\sigma}_{1,l}|a_1,...,a_l\right)\phi\left(\pmb{\sigma}_{l+1,n-1}|b_1,...,b_m\right) \right].\Label{Eq:I2}
\eea
The second line of \eqref{Eq:I1} (\eqref{Eq:I2}) can be verified by writing subcurrents of the form $\phi\left(...|b_1,...,b_m\right)$ ($\phi\left(...|a_1,...,a_l\right)$) according to the Berends-Giele recursion explicitly. Then $I_1$ and $I_2$  cancel out with the last two terms of $T_5$ anf $T_5'$ in (\ref{Eq:Summation}) and the RHS of \eqref{BTerm} turns into
\bea
\left[\phi\left(\pmb{\sigma}_{1,m}|b_1,...,b_m\right)\phi\left(\pmb{\sigma}_{m+1,n-1}|a_1,...,a_l\right)
-\phi\left(\pmb{\sigma}_{1,l}|a_1,...,a_l\right)\phi\left(\pmb{\sigma}_{l+1,n-1}|b_1,...,b_m\right) \right],\Label{Eq:RHSSimple}
\eea
which precisely matches with the expression (\ref{Eq:OffBCJRelation}) by considering simple chains $\mathcal{T}_{\pmb{A}}$ and $\mathcal{T}_{\pmb{B}}$. Note that the leftmost elements $x$ and $b$ in the original relation (\ref{Eq:OffBCJRelation}) can also be chosen as any other nodes $a_i$, $b_j$ in $\mathcal{T}_{\pmb{A}}$ and $\mathcal{T}_{\pmb{B}}$, in which the permutations $\{a_1,...,a_l\}$ and $\{b_1,...,b_m\}$ in  (\ref{Eq:RHSSimple}) are replaced by $\mathcal{T}_{\pmb{A}}|_{a_i}$ and $\mathcal{T}_{\pmb{B}}|_{b_i}$, respectively. This will be clear after we prove the relation based on graphs with branches.

\subsection{Graph-based BCJ relation with arbitrary $\mathcal{T}_{\pmb{A}}$ and $\mathcal{T}_{\pmb{B}}$}

Based on the relations established by simple chains, which have been proved in the previous section, we prove that the off-shell graph-based BCJ relation (\ref{Eq:OffBCJRelation}) holds when $\mathcal{T}_{\pmb{A}}$ and $\mathcal{T}_{\pmb{B}}$ refer to arbitrary connected tree graphs in this section. The main idea of the proof is to decompose both $\mathcal{T}_{\pmb{A}}$ and $\mathcal{T}_{\pmb{B}}$ into simple chains properly. Particularly, for an arbitrary tree structure $\mathcal{T}_{\pmb{B}}$, the summation over $\pmb{\beta}\in \mathcal{T}_{\pmb{B}}|_b$ in (\ref{Eq:DefofBr}) can be understood as a summation over simple chains $\mathcal{T}_{\pmb{B}_t}$ with nodes $\beta_1=b,\beta_2,...,\beta_m$ in turn,  because the leftmost element in the final permutations after shuffling nodes in both graphs together is always the element $b$. Therefore, the LHS of (\ref{Eq:OffBCJRelation}) can be displayed as
\bea
\Sl_{a\in \mathcal{T}_A}B^{(x)}(\pmb{\sigma}\,|\,\mathcal{T}_{\pmb {A}}|_a,\mathcal{T}_{\pmb{B}}|_b)=\Sl_{a\in \mathcal{T}_A}\biggl[\,\Sl_{t} B^{(x)}(\pmb{\sigma}\,|\,\mathcal{T}_{\pmb {A}}|_a,\mathcal{T}_{\pmb{B}_t}|_{b})\biggr],
\eea
where we summed over all possible simple chains $\mathcal{T}_{\pmb{B}_t}$ that are generated by the graph $\mathcal{T}_{\pmb{B}}$ when  $b$ is considered as the leftmost element.



\begin{figure}
\centering
  \includegraphics[width=0.8\textwidth]{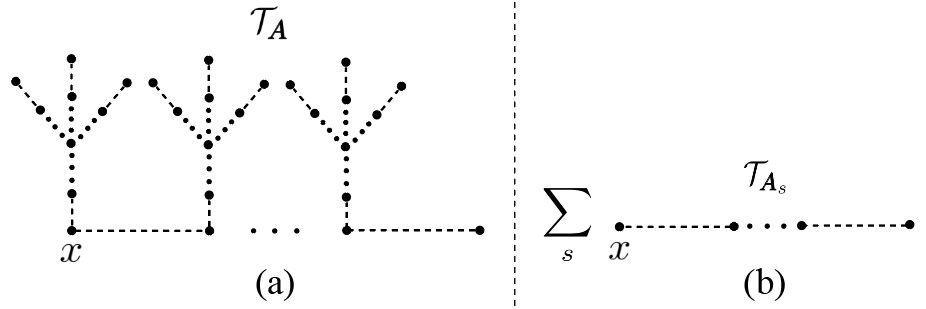}
  \caption{An arbitrary graph $\mathcal{T}_{\pmb{A}}$, as shown by (a), can be decomposed into a sum of simple chains $\mathcal{T}_{\pmb{A}_s}$ (shown by (b)) where node $x$ plays as an end node of each simple chain $\mathcal{T}_{\pmb{A}_s}$.}
  \label{Fig:TAdecomposition}
\end{figure}

The graph $\mathcal{T}_{\pmb{A}}$ can be decomposed into simple chains in a similar way, but this situation is much more subtle since the $\mathcal{T}_{\pmb{A}}|_a$ for distinct choices of $a$ correspond to different sets of permutations. Nevertheless, in \secref{sec:examples}, we have shown that the
star structure of $\mathcal{T}_{\pmb{A}}$ in \figref{Fig:StarGraphExample} (a) could be decomposed  into two simple chains $\mathcal{T}_{\pmb{A}_1}$ \figref{Fig:StarGraphExample} (c) and  $\mathcal{T}_{\pmb{A}_2}$ \figref{Fig:StarGraphExample} (d), by explicit evaluations. This provides a strong hint for studying the relations with general structures of $\mathcal{T}_{\pmb{A}}$: \emph{ One can pick out an arbitrary node $x\in \mathcal{T}_{\pmb{A}}$ and consider $x$ as the leftmost elements, then the allowed permutations $\pmb{\alpha}$ satisfy $\pmb{\alpha}\in\mathcal{T}_{\pmb{A}}|_x$. Such a given $\pmb{\alpha}$ defines a simple chain (denoted by $\mathcal{T}_{\pmb{A}_{s}}$), in which the node $x$ plays as one end of this chain, as shown by \figref{Fig:TAdecomposition}.} This decomposition can be understood as follows:
\begin{itemize}
\item According to the definition, permutations with $x$ as the leftmost node of $\mathcal{T}_{\pmb{A}}$ are just those permutations established by the simple chains $\mathcal{T}_{\pmb{A}_{s}}$ where the end node $x$ of $\mathcal{T}_{\pmb{A}_{s}}$ is regarded as the leftmost one, i.e.,
\bea
\mathcal{T}_{\pmb{A}}\big|_{x}\sim \Sl_s \mathcal{T}_{\pmb{A}_s}\big|_{x}. \Label{Eq:graphicBCJrelation6-0}
\eea
The symbol `$\sim$' means that when the node $x$ is considered as the leftmost one, the permutations established by $\mathcal{T}_{\pmb{A}}$ are same with those established by all simple chains $\mathcal{T}_{\pmb{A}_s}$.
\item We should  show that the permutations  $\mathcal{T}_{\pmb{A}}|_{a_r}$ for an arbitrary node $a_r\in\mathcal{T}_{\pmb{A}}$, $a_r\neq x$ can be decomposed as
\bea
(-)^{|a_r x|}\mathcal{T}_{\pmb{A}}\big|_{a_r}\sim\Sl_s (-)^{|a_r x|_s} \mathcal{T}_{\pmb{A}_s}\big|_{a_r}. \Label{Eq:graphicBCJrelation6}
\eea
In the above expression, the $|a_r x|_s$ denotes the distance between $a_r$ and $x$ in the corresponding simple chain $\mathcal{T}_{\pmb{A}_s}$. The above relation means that the summation over permutations of the graph $\mathcal{T}_{\pmb{A}}$ (with the sign $(-)^{|a_r x|}$) when $a_r$ is chosen as the leftmost element is equivalent to summing over all  $\mathcal{T}_{\pmb{A}_s}\big|_{a_r}$ (with the sign  $(-)^{|a_rx|_s}$) for all simple chains $\mathcal{T}_{\pmb{A}_s}$.

\end{itemize}

Having the decomposition (\ref{Eq:graphicBCJrelation6-0}) and the property (\ref{Eq:graphicBCJrelation6}), one can always express the LHS of \eqref{Eq:OffBCJRelation} as
\bea
\mathbb{B}^{(x)}(\pmb{\sigma}\,|\,\mathcal{T}_{\pmb {A}},\mathcal{T}_{\pmb{B}}|_b)&=&\Sl_{s}\Sl_{t}\biggl[\,\Sl_{a\in \mathcal{T}_{A_s}}B^{(x)}(\pmb{\sigma}\,|\,\mathcal{T}_{\pmb {A}_s}|_a,\mathcal{T}_{\pmb{B}_t}|_b)\biggr]\nn
&=&\Sl_{s}\Sl_{\pmb{\alpha}\in \mathcal{T}_{\pmb {A}_s}|_x}\Sl_{t}\Sl_{\pmb{\beta}\in \mathcal{T}_{\pmb{B}_t}|_b} \Big[\phi(\pmb{\sigma}_{1,i}|\pmb{\beta})\phi(\pmb{\sigma}_{i+1,l}|\pmb{\alpha}) -\phi(\pmb{\sigma}_{1,{l-i}}|\pmb{\alpha})\phi(\pmb{\sigma}_{l-i+1, l}|\pmb{\beta})\Big]\nn
&=&\Sl_{\pmb{\alpha}\in\mathcal{T}_{\pmb{A}}|_x}\Sl_{\pmb{\beta}\in \mathcal{T}_{\pmb{B}}|_b}\Big[\phi(\pmb{\sigma}_{1,i}|\pmb{\beta})\phi(\pmb{\sigma}_{i+1,l}|\pmb{\alpha}) -\phi(\pmb{\sigma}_{1,{l-i}}|\pmb{\alpha})\phi(\pmb{\sigma}_{l-i+1, l}|\pmb{\beta})\Big],
\eea
where the decompositions of $\mathcal{T}_{\pmb{A}}$ and $\mathcal{T}_{\pmb{B}}$ into simple chains, as well as the graph-based BCJ relation for simple chains which was proven in the previous subsection, have been applied. The summations $\Sl_{s}$ and $\Sl_{t}$ mean that we sum over all possible simple chains   $\mathcal{T}_{\pmb{A}_s}$ and $\mathcal{T}_{\pmb{B}_t}$ corresponding to  $\mathcal{T}_{\pmb{A}}$ and $\mathcal{T}_{\pmb{B}}$. Therefore, the proof of  (\ref{Eq:OffBCJRelation}) for arbitrary connected tree graphs  $\mathcal{T}_{\pmb{A}}$ and $\mathcal{T}_{\pmb{B}}$  have been completed. In the remaining of this section, we prove the crucial property (\ref{Eq:graphicBCJrelation6}).

\subsubsection{Proof of the property (\ref{Eq:graphicBCJrelation6})}

\begin{figure}
\centering
  \includegraphics[width=0.95\textwidth]{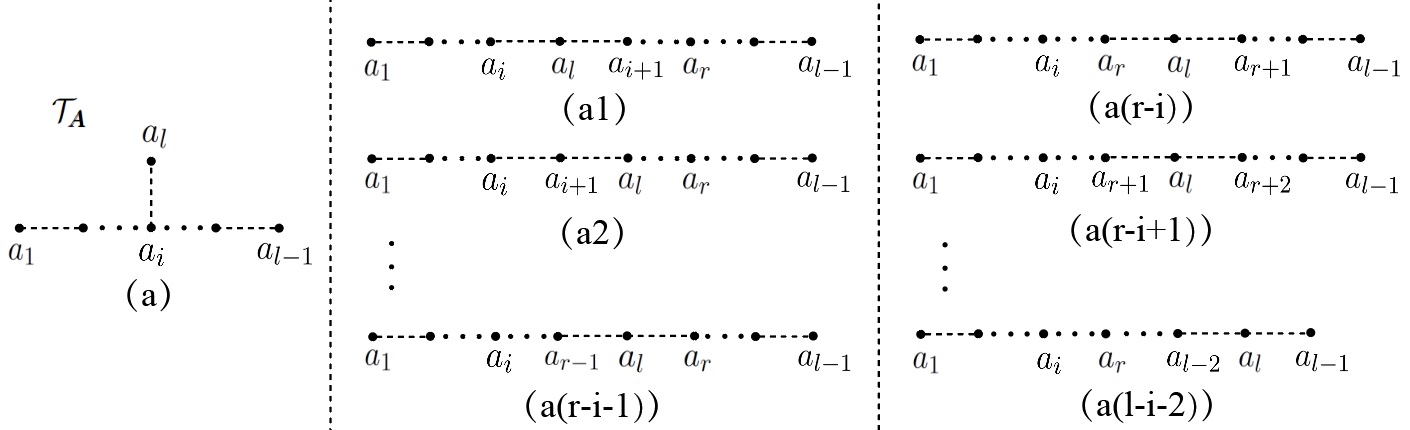}
  \caption{The graph (a) is a typical graph where a node $a_l$ is planted on $a_i$ ($i<r$) of a simple chain. The corresponding simple chains for (a) are given by (a1), ..., (a(r-i-1)) (where $a_l$ is located between $a_i$ and $a_r$) and  (a(r-i))... (a(l-i-2)) (where $a_l$ is inserted at the positions that occur after $a_r$ ).  }
  \label{Fig:any_alpha_decomposition1}
\end{figure}
We prove the decomposition property  (\ref{Eq:graphicBCJrelation6}) by the following steps.

\noindent{\bf Step-1}~~ We first consider the decomposition of the graph \figref{Fig:any_alpha_decomposition1} (a) where a single node $a_l$ is planted at the node $a_i$ for a given $i$ ($1<i<l-1$). Without loss of generality, we choose $x=a_1$ in the relation (\ref{Eq:OffBCJRelation}), then the simple chains $\mathcal{T}_{\pmb{A}_s}$ in (\ref{Eq:graphicBCJrelation6-0}) are based on the permutations $\pmb{\alpha}\in\{a_1,...,a_i,a_l\shuffle\{a_{i+1},...,a_{l-1}\}\}$. In the following, we prove the decomposition (\ref{Eq:graphicBCJrelation6}) for $r=1,2,...,i$ and $r=i+1,...,l-1$, separately. The case $r=l$ is included as a special situation of the step-2.
\begin{itemize}
\item (i). If $r=1,2,...,i$, permutations $\mathcal{T}_{\pmb{A}}\big|_{a_r}$ and the sign $(-)^{|a_ra_1|}$ are given by
\bea
(-1)^{r-1}\big\{a_{r-1},...,a_1\big\}\shuffle\Big\{a_r,\big\{a_{r+1}...,a_i,\{a_l\}\shuffle\{a_{i+1},...,a_{l-1}\}\big\}\Big\}
\eea
that are just those permutations $\mathcal{T}_{\pmb{A}_s}\big|_{a_r}$ established by the simple chains $\mathcal{T}_{\pmb{A}_s}$ with the correct sign.

\item (ii). If $r=i+1,...,l-1$, the LHS of (\ref{Eq:graphicBCJrelation6}) becomes
\bea
(-)^{r-1}\Bigl\{a_r,\pmb{A}_{r+1,l-1}\shuffle\bigl\{\pmb{A}^T_{i,r-1}, \pmb{A}^T_{1,i-1}\shuffle\{a_l\}\bigr\}  \Bigr\}, \Label{Eq:graphicBCJrelation7}
\eea
where $\pmb{A}_{i,j}$ denotes the permutation $\{a_i,a_{i+1},...,a_j\}$, and $\pmb{A}^T_{i,j}$ refers to the inverse of $\pmb{A}_{i,j}$.
On the other hand, the permutations ${T}_{\pmb{A}_s}|_{a_r}$ corresponding to the simple chains \figref{Fig:any_alpha_decomposition1} (a1), ... , (a(r-i-1)),  where $a_l$ is located between $a_i$ and $a_r$, have the form
\bea
(-)^{r}\Bigl\{a_r,\pmb{A}_{r+1,l-1}\shuffle\bigl\{ \pmb{A}^T_{i+1,r-1}\shuffle\{a_l\},\pmb{A}^T_{1,i}\bigr\} \Bigr\}, \Label{Eq:graphicBCJrelation9}
\eea
while those ${T}_{\pmb{A}_s}|_{a_r}$ established by the simple chains \figref{Fig:any_alpha_decomposition1} (a(r-i)), ..., (a(l-i-2)) are collected together as
\bea
&&(-)^{r-1}\Bigl\{a_r,\pmb{A}_{r+1,l-1}\shuffle \pmb{A}^T_{1,r-1}\shuffle\{a_l\}\Bigr\}. \Label{Eq:graphicBCJrelation8}
\eea
Since $\pmb{A}^T_{1,r-1}=\bigl\{\pmb{A}^T_{i+1,r-1},\pmb{A}^T_{1,i}\bigr\}$ and the permutations (\ref{Eq:graphicBCJrelation9}), (\ref{Eq:graphicBCJrelation8}) have opposite signs, the summation over all permutations $\pmb{\alpha}\in\mathcal{T}_{\pmb{A}_s}\big|_{a_r}$ then turns into a summation over permutations (\ref{Eq:graphicBCJrelation7}). Thus the property (\ref{Eq:graphicBCJrelation6}) holds for this case.

\end{itemize}

\begin{figure}
  \centering
  \includegraphics[width=0.75\textwidth]{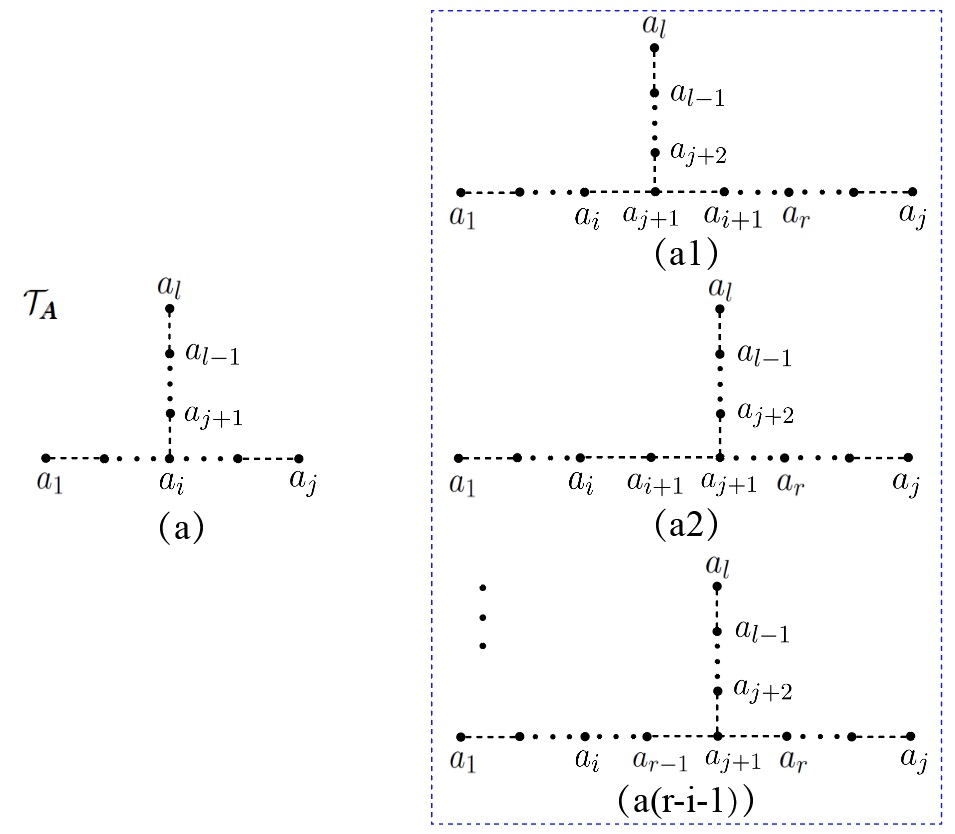}
  \caption{ Graph (a) is constructed by planting a simple chain with nodes $\{a_{j+1},...,a_l\}$ at $a_i$. According to the inductive assumption, the simple chains corresponding to permutations  (\ref{Eq:p1}), (\ref{Eq:p2}), ..., (\ref{Eq:p(r-i-1)}) can be reexpressed by  (a1),...,(a(r-i-1)), respectively. }
  \label{Fig:any_alpha_decomposition2}
\end{figure}
\noindent{\bf Step-2}~~ When the node $a_l$ in  \figref{Fig:any_alpha_decomposition1} (a) is replaced by a simple chain, we get a more complicated graph \figref{Fig:any_alpha_decomposition2} (a). Permutations established by \figref{Fig:any_alpha_decomposition2} (a) when $a_1$ is considered as the leftmost element are given by $\pmb{\alpha}\in \{\pmb{A}_{1,i},\pmb{A}_{i+1,j}\shuffle \pmb{A}_{j+1,l}\}$. Therefore, a given permutation $\pmb{\alpha}=\{\alpha_1,...,\alpha_l\}$ defines a simple chain $\mathcal{T}_{\pmb{A}_s}$ with nodes $\alpha_1,...,\alpha_l$ in turn. To prove the decomposition (\ref{Eq:graphicBCJrelation6}), we again consider the cases $r=1,...,i$ and $r=i=1,...,j$ separately. Since the two branches attached to the node $a_i$ have the same status, the cases $r=j+1,...,l$ are automatically proven once (\ref{Eq:graphicBCJrelation6}) has been proven for  $r=i+1,...,j$.

When $r=1,...,i$, the decomposition (\ref{Eq:graphicBCJrelation6}) can be directly verified, via replacing the node $a_l$ in the case (i) of step-1 by a simple chain. The cases with $r=i+1,...,j$ can be proved inductively, under the assumption that (\ref{Eq:graphicBCJrelation6}) already holds for the graphs where a shorter simple chain is planted at the node $a_i$. The starting point of this proof is the graph  \figref{Fig:any_alpha_decomposition1} (a) which has been studied in step-1.
To proceed, we write the LHS of (\ref{Eq:graphicBCJrelation6}) explicitly
\bea
(-)^{|a_ra_1|}\mathcal{T}_{\pmb{A}}\big|_{a_r}
=(-)^{r-1}\Bigl\{a_r,\pmb{A}_{r+1,j}\shuffle\bigl\{\pmb{A}^T_{i,r-1}, \pmb{A}^T_{1,i-1}\shuffle\pmb{A}_{j+1,l}\bigr\} \Bigr\}. \Label{Eq:graphicBCJrelation10}
\eea
On the other hand, the simple chains $\mathcal{T}_{\pmb{A}_s}$ that are defined by $\mathcal{T}_{\pmb{A}}$ can be classified according to different positions of the node $a_{j+1}$ in the corresponding permutations $\pmb{\alpha}\in \mathcal{T}_{\pmb{A}}|_{a_1}=\left\{\pmb{A}_{1,i},\pmb{A}_{i+1,j}\shuffle \pmb{A}_{j+1,l}\right\}$:
\bea
&\{\pmb{A}_{1,i},a_{j+1},\pmb{A}_{i+1,j}\shuffle\pmb{A}_{j+2,l}\},&\Label{Eq:p1}\\
&\{\pmb{A}_{1,i+1},a_{j+1},\pmb{A}_{i+2,j}\shuffle\pmb{A}_{j+2,l}\},&\Label{Eq:p2}\\
&\vdots&\nn
&\{\pmb{A}_{1,r-1},a_{j+1},\pmb{A}_{r,l-1}\shuffle\pmb{A}_{j+2,l}\},&\Label{Eq:p(r-i-1)}\\
&\{\pmb{A}_{1,r},\pmb{A}_{r+1,l-1}\shuffle\pmb{A}_{j+1,l}\}.&\Label{Eq:p(r-i)}
\eea
Permutations (\ref{Eq:p1}), (\ref{Eq:p2}), ..., (\ref{Eq:p(r-i-1)}) can be regarded as the permutations corresponding to graphs \figref{Fig:any_alpha_decomposition1} (a1), (a2), ..., (a(r-i-1))  when $a_1$ is considered as the leftmost element,  while (\ref{Eq:p(r-i)}) is the collection of all the permutations when $a_{j+1}$ is located on the RHS of $a_r$. \emph{By the help of the inductive assumption, the permutations $\mathcal{T}_{\pmb{A}_s}\big|_{a_r}$, where $\mathcal{T}_{\pmb{A}_s}$ corresponding to  (\ref{Eq:p1}), (\ref{Eq:p2}), ..., (\ref{Eq:p(r-i-1)}) can be rewritten as a summation over permutations established  by graphs \figref{Fig:any_alpha_decomposition2} (a1), (a2), ..., (a(r-i-1)), when $a_r$ is considered as the leftmost element:}
\bea
&&(-)^{r}\Bigl\{a_r,\pmb{A}_{r+1,j}\shuffle\Bigl(\bigl\{ \pmb{A}^T_{i+1,r-1},a_{j+1},\pmb{A}_{j+2,l}\shuffle\pmb{A}^T_{1,i} \bigr\} \nn
&&~~~~~~~~~~~~~~~~~~~~~~\,+\bigl\{\pmb{A}^T_{i+2,r-1},a_{j+1}, \pmb{A}_{j+2,l}\shuffle\pmb{A}^T_{1,i+1}\bigr\}+\cdots
+\bigl\{a_{j+1},\pmb{A}_{j+2,l}\shuffle \pmb{A}^T_{1,r-1}\bigr\}\Bigl) \Bigr\}. \Label{Eq:graphicBCJrelation12}
\eea
In addition, all the simple chains corresponding to (\ref{Eq:p(r-i)}) together establish the following permutations
\bea
&&(-)^{r-1}\Bigl\{a_r,\pmb{A}_{r+1,j}\shuffle \Bigl(\pmb{A}_{1,r-1}^T\shuffle\pmb{A}_{j+1,l}\Bigr)\Bigr\}. \Label{Eq:graphicBCJrelation11}
\eea
Noting that the permutations in the parentheses of the above expression can be decomposed into the permutations in the  parentheses in (\ref{Eq:graphicBCJrelation12}) (with an opposite sign) and the additional ones
\bea
\bigl\{\pmb{A}^T_{i,r-1}, \pmb{A}^T_{1,i-1}\shuffle\pmb{A}_{j+1,l}\bigr\},
\eea
the overall contributions of (\ref{Eq:graphicBCJrelation12}) and (\ref{Eq:graphicBCJrelation11}) precisely match  with (\ref{Eq:graphicBCJrelation9}) after cancellation. Hence (\ref{Eq:graphicBCJrelation6}) for $r=i+1,...,j$  is proven.

\begin{figure}
\centering
  \includegraphics[width=0.8\textwidth]{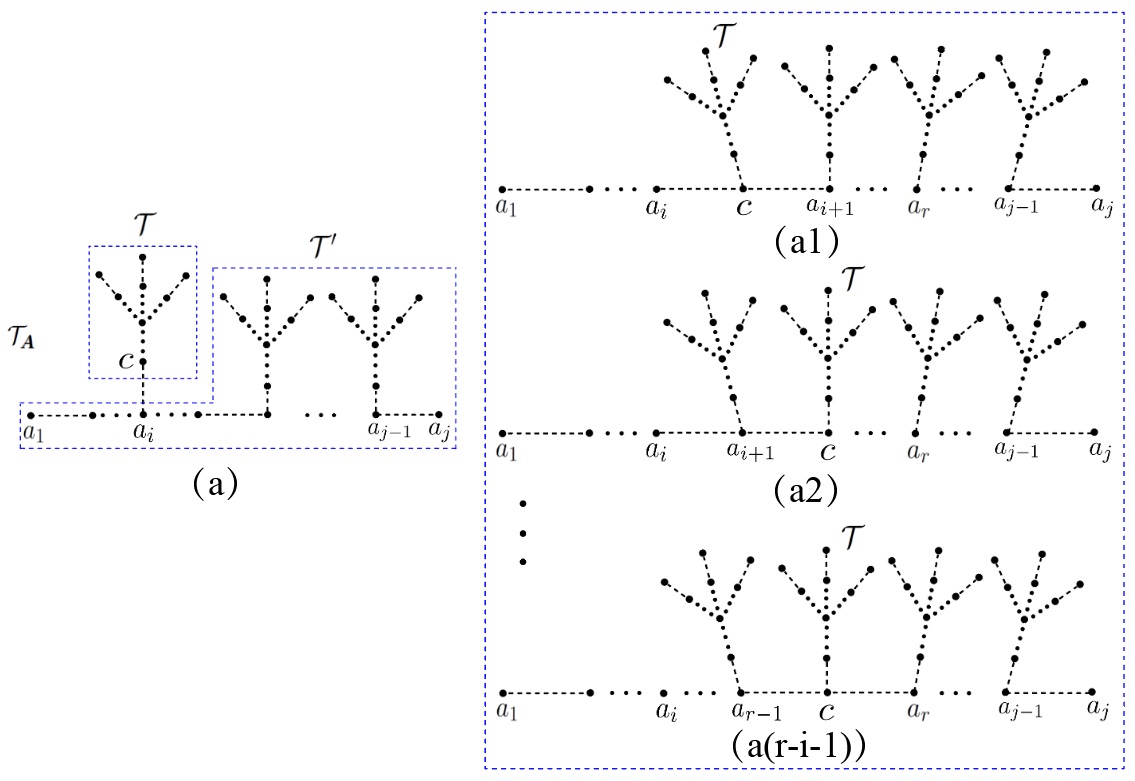}
  \caption{ Graph $\mathcal{T}_{\pmb{A}}$ in (a) can be separated into two subtree sturctures $\mathcal{T}$ and $\mathcal{T}'$. Graphs (a1), ..., (a(r-i-1)) denote the contributions of the sets of simple  chains that are corresponding to  (\ref{Eq:genp1}), (\ref{Eq:genp2}), ... and (\ref{Eq:genpr-i-1}).  }
  \label{Fig:any_alpha_decomposition3}
\end{figure}

\noindent{\bf Step-3}~~The inductive proof in step-2 can be generalized into an arbitrary connected tree graph  $\mathcal{T}_{\pmb{A}}$ (see \figref{Fig:any_alpha_decomposition3} (a)), straightforwardly.
In general, a tree graph can be described by attaching tree structures to the path from $a_1$ to $a_j$ as shown by \figref{Fig:any_alpha_decomposition3} (a). Without loss of generality, we assume that the subtree structures are planted at the nodes belonging to $\{a_{i+1},...,a_{j-1}\}$. Supposing that the first subtree structure $\mathcal{T}$ is attached to the node $a_i$ and the nearest-to-$a_i$ node in $\mathcal{T}$ is $c$, the full tree $\mathcal{T}_{\pmb{A}}$ then consists of two subtrees  $\mathcal{T}$ and  $\mathcal{T}'$, as shown by graph \figref{Fig:any_alpha_decomposition3} (a). We have already demonstrated in step-1 and step-2 that the decomposition (\ref{Eq:graphicBCJrelation6}) is naturally satisfied in the case of $1<r\leq i$. When $i<r\leq j$, the LHS of (\ref{Eq:graphicBCJrelation6}) is given by
\bea
(-1)^{r-1}(\mathcal{T}'|_{a_r}\shuffle \mathcal{T}|_{c})\big|_{a_i\prec c}.\Label{Eq:graphicBCJrelation13}
\eea
Meanwhile, the simple chains  $\mathcal{T}_{\pmb{A}_s}$ which correspond to the permutations $\pmb{\alpha}\in \mathcal{T}_{\pmb{A}}\big|_{a_1}$ can be classified
according to different positions of $c$ in $\pmb{\alpha}\in \mathcal{T}_{\pmb{A}}\big|_{a_1}$:
\bea
&&(\mathcal{T}'|_{a_1}\shuffle \mathcal{T}|_{c})\big|_{a_{i}\prec c\prec a_{i+1}},\Label{Eq:genp1}\\
&&(\mathcal{T}'|_{a_1}\shuffle \mathcal{T}|_{c})\big|_{a_{i+1}\prec c\prec a_{i+2}},\Label{Eq:genp2}\\
&&~~~~~~~~~~~~~~\vdots\nn
&&(\mathcal{T}'|_{a_1}\shuffle \mathcal{T}|_{c})\big|_{a_{r-1}\prec c\prec a_{r}},\Label{Eq:genpr-i-1}\\
&&(\mathcal{T}'|_{a_1}\shuffle \mathcal{T}|_{c})\big|_{a_{r}\prec c}.\Label{Eq:genpr-i}
\eea
Those simple chains corresponding to (\ref{Eq:genp1}), (\ref{Eq:genp2}), ... and (\ref{Eq:genpr-i-1})  can convert into the graphs \figref{Fig:any_alpha_decomposition3} (a1), ...,(a(r-i-1)) under the inductive assumption that (\ref{Eq:graphicBCJrelation6}) holds if $\mathcal{T}$ is replaced by a lower-tree structure. Therefore, the permutations established by these simple chains when $a_r$ is considered as the leftmost elements are
\bea
&&(-1)^r(\mathcal{T}'|_{a_r}\shuffle \mathcal{T}|_{c})\big|_{a_{i+1}\prec c\prec a_{i}},\Label{Eq:Genp1}\\
&&(-1)^r(\mathcal{T}'|_{a_r}\shuffle \mathcal{T}|_{c})\big|_{a_{i+2}\prec c\prec a_{i+1}},\Label{Eq:Genp2}\\
&&~~~~~~~~~~~~~~\vdots\nn
&&(-1)^r(\mathcal{T}'|_{a_r}\shuffle \mathcal{T}|_{c})\big|_{a_{r}\prec c\prec a_{r-1}},\Label{Eq:Genpr-i-1}
\eea
where we should note that the relative order of elements $a_1,...,a_r$ has been reversed when $a_r$ is considered as the leftmost one. The permutations corresponding to the simple chains that refer to (\ref{Eq:genpr-i}), when $a_r$ is the leftmost element, are presented by
\bea
(-1)^{r-1}(\mathcal{T}'|_{a_r}\shuffle \mathcal{T}|_{c})\big|_{a_{r}\prec c}.\Label{Eq:Genpr-i}
\eea
After cancellation, the overall contributions of (\ref{Eq:Genp1}), (\ref{Eq:Genp2}), ..., (\ref{Eq:Genpr-i-1}) and (\ref{Eq:Genpr-i}) precisely match with (\ref{Eq:graphicBCJrelation13}). Hence the proof of (\ref{Eq:graphicBCJrelation6}) is completed.

\section{Further discussions}\label{sec:FurtherDiscussions}

Having proven the graph-based BCJ relation  (\ref{Eq:OffBCJRelation}) for the Berends-Giele currents in BS, we now provide more discussions including the on-shell limits, the antisymmetry under the exchanging $\mathcal{T}_{\pmb{A}}\leftrightarrow\mathcal{T}_{\pmb{B}}$, as well as the implication in YM theory.

%

\subsection{On-shell limit}

Supposing that there are $n-1$ external on-shell particles, the on-shell limit of the LHS of (\ref{Eq:OffBCJRelation}) is taken as follows
\bea
\lim_{s_{1\dots n-1}\rightarrow0}s_{1\dots n-1}\mathbb{B}^{(x)}(\pmb{\sigma}\,|\,\mathcal{T}_{\pmb {A}},\mathcal{T}_{\pmb{B}}|_b)
&=&\lim_{s_{1\dots n-1}\rightarrow0}\Sl_{a\in\mathcal{T}_{\pmb{A}}}s_{1\dots n-1}\mathbb{B}^{(x)}(\pmb{\sigma}\,|\,\mathcal{T}_{\pmb {A}}|_a,\mathcal{T}_{\pmb{B}}|_b) \nn
&=&\Sl_{a\in\mathcal{T}_{\pmb{A}}}\Sl_{c\in\mathcal{T}_{\pmb{B}}}\Sl_{\pmb{\alpha}\in\mathcal{T}_{\pmb{A}}|_a }\Sl_{\pmb{\beta}\in\mathcal{T}_{\pmb{B}}|_b } (-1)^{|xa|}s_{ac}\,\mathcal{A}\bigl(\pmb{\sigma}_{1,n}\big|\,\pmb{\alpha}\shuffle\pmb{\beta},\sigma_n\bigr),
\eea
where the relation (\ref{Eq:BG-Amp}) between Berends-Giele currents and amplitudes has been applied and the sign for the node $x$ ($x\in \mathcal{T}_{\pmb{A}}$) is chosen as $+1$. The RHS of (\ref{Eq:OffBCJRelation}), when multiplied by $s_{1\dots n-1}$ turns into zero in the limit  $s_{1\dots n-1}\to 0$ because it does not involve   $\frac{1}{s_{1\dots n-1}}$. The on-shell limit hence provides the corresponding relation for on-shell BS amplitudes
\bea
\Sl_{a\in\mathcal{T}_{\pmb{A}}}\Sl_{c\in\mathcal{T}_{\pmb{B}}}\Sl_{\pmb{\alpha}\in\mathcal{T}_{\pmb{A}}|_a }\Sl_{\pmb{\beta}\in\mathcal{T}_{\pmb{B}}|_b } (-1)^{|xa|}s_{ac}\, \mathcal{A}\bigl(\pmb{\sigma}_{1,n}\big|\,\pmb{\alpha}\shuffle\pmb{\beta} |_{c\prec a},\sigma_n\bigr)=0, \Label{Eq:OnShellRelations}
\eea
whose Yang-Mills version was first proposed in  \cite{Hou:2018bwm}. Here, we emphasize that the above on-shell relation can be understood as a proper combination of the general BCJ relations \cite{BjerrumBohr:2009rd,Stieberger:2009hq,Chen:2011jxa}. This fact has  already been proven in \cite{Hou:2018bwm}. Furthermore, the general BCJ relations can be generated by the fundamental ones \cite{Feng:2010my,Ma:2011um}. In this sense, the relation  (\ref{Eq:OnShellRelations}) based on graphs is essentially a proper combination of the fundamental BCJ relations, and cannot further reduce the number of independent color-ordered amplitudes.

\subsection{The antisymmetry under the exchanging of $\mathcal{T}_{\pmb{A}}$ and $\mathcal{T}_{\pmb{B}}$}

\begin{figure}
\centering
  \includegraphics[width=0.7\textwidth]{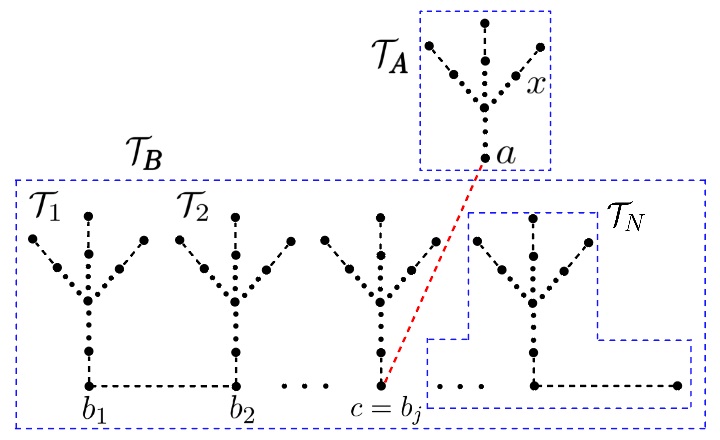}
  \caption{ This graph is constructed by connecting two arbitrary nodes $a$ and $c$ in graphs $\mathcal{T}_{\pmb{A}}$ and $\mathcal{T}_{\pmb{B}}$ together. The nodes $x$ and $b=b_1$ are taken as the leftmost elements in $\mathcal{T}_{\pmb{A}}$ and $\mathcal{T}_{\pmb{B}}$, respectively. }
  \label{Fig:GenStructure}
\end{figure}

When we exchange the roles of the two graphs  $\mathcal{T}_{\pmb{A}}$ and $\mathcal{T}_{\pmb{B}}$ and the roles of the two nodes $x\in \mathcal{T}_{\pmb{A}}$ and $b\in \mathcal{T}_{\pmb{B}}$, the RHS of the relation (\ref{Eq:OffBCJRelation}) is only changed by a minus sign. However, this antisymmetry is not transparent for the LHS of the relation (\ref{Eq:OffBCJRelation}). To convert the LHS of  (\ref{Eq:OffBCJRelation}) into an expression manifesting the antisymmetry, we notice that a general graph of $\mathcal{T}_{\pmb{A}}$ and $\mathcal{T}_{\pmb{B}}$ can be shown by \figref{Fig:GenStructure}, which corresponds to the contribution with a given $a\in \mathcal{T}_{\pmb{A}}$ and $c\in \mathcal{T}_{\pmb{B}}$ in \eqref{Eq:DefofBr}. Here, we have  connected a line between $a$ and $c$ to reflect the relative order between them. KK relation allows one to reverse the orders of elements in the trees $ \mathcal{T}_{1}$, ..., $ \mathcal{T}_{N}$ in turn, with an overall factor $(-1)^{m-j}$ ($m$ is the total number of nodes in  $\mathcal{T}_{\pmb{B}}$). After this manipulation, the permutation $\mathcal{T}_{\pmb{B}}|_{b}$  is converted into $(\mathcal{T}_{\pmb{B}}|_{c})^T$. Therefore, the LHS of (\ref{Eq:OffBCJRelation}) turns into
\bea
&&\Sl_{c\in \mathcal{T}_{\pmb{B}}}\Sl_{a\in \mathcal{T}_{\pmb{A}}}(-1)^{m-|bx|}\Sl_{\shuffle} s_{ac}\,\phi\left(\,\pmb{\sigma}\,|\,(\mathcal{T}_{\pmb{B}}|_{c})^T,\mathcal{T}_{\pmb{A}}|_{a}\,\right)\nn
&&~~~~~~~~~~~~~~~~~~~=\Sl_{\pmb{\alpha}\in { \mathcal{T}_{\pmb{A}}|_x}}\Sl_{\pmb{\beta}\in \mathcal{T}_{\pmb{B}}|_b} \Big[\phi(\pmb{\sigma}_{1,i}|\pmb{\beta})\phi(\pmb{\sigma}_{i+1,l}|\pmb{\alpha}) -\phi(\pmb{\sigma}_{1,{l-i}}|\pmb{\alpha})\phi(\pmb{\sigma}_{l-i+1, l}|\pmb{\beta})\Big]. \Label{Eq:AntiSymOffBCJRelation}
\eea
In the above relation, the overall sign was evaluated as $(-1)^{m-j}(-1)^{|ax|}=(-1)^{m-(j+|ax|)}=(-1)^{m-(|bx|)}$, where we denote the distance (the number of edges) between $a$ and $x$ in the graph $\mathcal{T}_{\pmb{A}}$  by $|ax|$. The distance between $b$ and $x$, when $a$ and $c$ are connected by an edge, is denoted by $|bx|$. When the reflection relation (\ref{Eq:BsProperty2}) is applied, the antisymmetry of the LHS under the exchanging  $\mathcal{T}_{\pmb{A}}\leftrightarrow\mathcal{T}_{\pmb{B}}$ and $x\leftrightarrow b$ is then manifest:
\bea
&&\Sl_{c\in \mathcal{T}_{\pmb{B}}}\Sl_{a\in \mathcal{T}_{\pmb{A}}}(-1)^{l-|xb|}\Sl_{\shuffle} s_{ac}\,\phi\left(\,\pmb{\sigma}\,|\,(\mathcal{T}_{\pmb{A}}|_{a})^T,\mathcal{T}_{\pmb{B}}|_{c}\,\right)\nn
&=&(-1)\Sl_{c\in \mathcal{T}_{\pmb{B}}}\Sl_{a\in \mathcal{T}_{\pmb{A}}}(-1)^{m-|bx|}\Sl_{\shuffle} s_{ac}\,\phi\left(\,\pmb{\sigma}\,|\,(\mathcal{T}_{\pmb{B}}|_{c})^T,\mathcal{T}_{\pmb{A}}|_{a}\,\right),
\eea
where $l$ denotes the number of elements in $\mathcal{T}_{\pmb{A}}$. If both  $\mathcal{T}_{\pmb{A}}$ and $\mathcal{T}_{\pmb{B}}$ are simple chains which respectively involve $x$ and $b$ as an end node, the antisymmetric formula (\ref{Eq:AntiSymOffBCJRelation}) becomes the relation which was established via free Lie algebras  \cite{Frost:2020bmk,Frost:2021asr}.

\subsection{Graph-based BCJ relation for Berends-Giele currents in YM}

At on-shell level, the graph-based BCJ relation (\ref{Eq:OnShellRelations}) can be straightforwardly extended to the partial amplitudes in other theories such as YM or nonlinear sigma model, through the framework of Cachazo-He-Yuan formula \cite{Cachazo:2013gna,Cachazo:2013hca,Cachazo:2013iea,Cachazo:2014nsa}. In the following, we extend the off-shell relation (\ref{Eq:OffBCJRelation}) to YM theory, by the help of the decomposition formula proposed in \cite{Wu:2021bcy}.

The Berends-Giele currents with Feynman gauge in YM  have been shown to satisfy a decomposition formula \cite{Wu:2021bcy} which consists of a part vanishing under on-shell limit and an effective current $\W J^{\,\rho}(1,2,...,n-1)$ satisfying the following decomposition formula
\bea
\W J^{\rho}(1,2,\dots,n-1)=\Sl_{\pmb{\sigma}\in P(2,n-1)}\,N_A^{\rho}(1,\pmb{\sigma})\phi(1,2,\dots,n-1 | 1,\pmb{\sigma}).\Label{Eq:GenForm1}
\eea
Here, the off-shell polynomial numerators $N_A^{\rho}(1,\pmb{\sigma})$ (which was constructed by a graphic rule in \cite{Wu:2021bcy}),  can generate the BCJ numerators with Lie symmetries. As pointed in \cite{Wu:2021bcy}, the $N_A^{\rho}(1,\pmb{\sigma})$ satisfies the following factorization property
\bea
N_{A}^\rho(1,\pmb{\sigma})
&=&N_{A}^\nu(1,\pmb{\sigma}_L)\Big[N_{C}^{\nu\rho}(\pmb{\sigma}_R) -\big(N_B(\pmb{\sigma}_R)\cdot2k_{1,\pmb{\sigma}_L}\big)\eta^{\nu\rho}\Big], \Label{Eq:DecOffshellNumerators}
\eea
where $N_{C}^{\nu\rho}(\pmb{\sigma}_R)$ and $N_B^{\mu}(\pmb{\sigma}_R)$ are other two types of numerators, the permutations on both sides are related by $\pmb{\sigma}=\{\pmb{\sigma}_L,\pmb{\sigma}_R\}$. Having  \eqref{Eq:GenForm1} and \eqref{Eq:DecOffshellNumerators} in hands, we dress the numerators $N_A^{\rho}(1,\pmb{\sigma})$ to the Berends-Giele currents of BS in (\ref{Eq:DefofBr}), then obtain

\bea
B_{\text{YM}}^{(x)}\big(\mathcal{T}_{\pmb {A}}|_a,\mathcal{T}_{\pmb{B}}|_1 \big)
&\equiv&\Sl_{\pmb{\sigma}\in P(2,n-1)}\,N_A^{\rho}(1,\pmb{\sigma})\,B^{(x)}\big(1,\pmb{\sigma}\,\big|\,\mathcal{T}_{\pmb {A}}|_a,\mathcal{T}_{\pmb{B}}|_1\big)\nn
&=&(-)^{|ax|}\Sl_{c\in \mathcal{T}_{\pmb{B}}}\Sl_{\pmb{\alpha}\in \mathcal{T}_{\pmb{A}}|_a}\Sl_{\pmb{\beta}\in \mathcal{T}_{\pmb{B}}|_1} \left[\,\Sl_{\pmb{\gamma}\in \pmb{\alpha}\shuffle\pmb{\beta}|_{c\prec a}}s_{ac}\, \W J^\rho(\pmb{\gamma})\,\right].
\eea
where we have set the $b$ in (\ref{Eq:DefofBr}) as the particle $1$. When all $a\in\mathcal{T}_{\pmb{A}}$ are summed over, \eqref{Eq:OffBCJRelation} implies
\bea
\mathbb{B}_{\text{YM}}^{(x)}\big(\mathcal{T}_{\pmb {A}},\mathcal{T}_{\pmb{B}}|_1\big)
&\equiv&\Sl_{a\in \mathcal{T}_A}B_{\text{YM}}^{(x)}\big(\mathcal{T}_{\pmb {A}}|_a,\mathcal{T}_{\pmb{B}}|_1\bigr) \nn
&=&\Sl_{\pmb{\sigma}\in P(2,n-1)}N^\rho_{A}(1,\pmb{\sigma})\bigg[\Sl_{\pmb{\alpha}\in \mathcal{T}_{\pmb{A}}|_a}\Sl_{\pmb{\beta}\in \mathcal{T}_{\pmb{B}}|_1} \Big(\phi(1,\pmb{\sigma}_{2,i}|\pmb{\beta})\phi(\pmb{\sigma}_{i+1,l}|\pmb{\alpha}) -\phi(1,\pmb{\sigma}_{2,{l-i}}|\pmb{\alpha})\phi(\pmb{\sigma}_{l-i+1, l}|\pmb{\beta})\Big)\bigg] \nn
&=&\Sl_{\pmb{\sigma}\in P(2,n-1)}N^{\rho}_{A}(1,\pmb{\sigma})\Sl_{\pmb{\alpha}\in\mathcal{T}_{\pmb{A}}|_a}\Sl_{\pmb{\beta}\in \mathcal{T}_{\pmb{B}}|_1} \phi(1,\pmb{\sigma}_{2,i}|\pmb{\beta}) \phi(\pmb{\sigma}_{i+1,l}|\pmb{\alpha}). \Label{Eq:YMOoffShellBCJ}
\eea
Since only the $\mathcal{T}_{\pmb{B}}$ involves the node $1$, the second term on the second line of the above expression must vanish. When inserting (\ref{Eq:DecOffshellNumerators}) into \eqref{Eq:YMOoffShellBCJ}, we finally get the graph-based BCJ relation for the effective current $\W J^{\rho}(1,2,\dots,n-1)$ in YM
\bea
\mathbb{B}_{\text{YM}}^{(x)}(\mathcal{T}_{\pmb {A}},\mathcal{T}_{\pmb{B}}|_1)
&=&\Biggl[\,\Sl_{\pmb{\sigma}_L\in P(2,i)}N_{A}^\nu(1,\pmb{\sigma}_L)\phi\left(1,\pmb{\sigma}_L| \mathcal{T}_{\pmb{B}}|_1\right)\Biggr] \nn
&&~~~~~~\times\Biggl[\,\Sl_{\pmb{\sigma}_R\in P(i+1,n-1)}\left[N_{C}(\pmb{\sigma}_R) -N_{B}(\pmb{\sigma}_R)\cdot2k_{1,i} \right]^{\nu\rho}\phi\left(\pmb{\sigma}_R|\mathcal{T}_{\pmb{A}}\bigr|_x\right)\Biggr].
\eea
where $\pmb{\sigma}_L=\{\sigma_2,...,\sigma_i\}$ and $\pmb{\sigma}_R=\{\sigma_{i+1},...,\sigma_{n-1}\}$. In the on-shell limit, the RHS has to vanish when multiplied by $s_{1,...,n-1}$ ($s_{1,...,n-1}=k_n^2\to 0$). Hence this relation precisely matches with the corresponding relation  \cite{Hou:2018bwm} for on-shell YM amplitudes. An interesting result is that the first factor is the effective current (\ref{Eq:GenForm1}) with fewer external particles, while the second factor can be expressed via effective currents and the generalized strength tensors $F^{\mu\nu}\equiv k^{\mu}_{1,i}\widetilde{J}^{\nu}(1,...,i)- k^{\nu}_{1,i}\widetilde{J}^{\mu}(1,...,i)$ \cite{Wu:2021bcy}. Thus RHS of the above relation in YM thery can be expressed via lower-point effective currents.

\section{Conclusions} \label{sec:Conclusions}

In this work, the graph-based BCJ relation for the Berends-Giele currents in BS theory and YM theory, was systematically studied. In the relation, distinct permutations of external particles are characterized by graphs, while any connected tree graph is shown to be equivalent to a proper combination of simple chains. The RHS of the relation (\ref{Eq:OffBCJRelation}) in BS is given by the difference between products of two subcurrents. When both sides are dressed by the off-shell numerators, the relation (\ref{Eq:OffBCJRelation}) turns into the off-shell graph-based BCJ relation in YM theory. The latter reproduces the relation observed in earlier work \cite{Hou:2018bwm} in the on-shell limit. When the graphs are simple chains and the relation is transformed into the antisymmetric form  (\ref{Eq:AntiSymOffBCJRelation}),
we get the one \cite{Frost:2020bmk,Frost:2021asr} which was derived by free Lie algebra.

\section*{Acknowledgments}
We would like to thank Shiquan Ma, Xinhai Tong and Chongsi Xie for helpful discussions. This work is supported by NSFC under Grant No. 11875206, Jiangsu Ministry of Science and Technology under contract BK20170410.


\appendix

\bibliographystyle{JHEP}
\bibliography{GraphBCJ}

\providecommand{\href}[2]{#2}\begingroup\raggedright\begin{thebibliography}{10}

\bibitem{Stieberger:2016lng}
S.~Stieberger and T.~R. Taylor, {\it {New relations for Einstein-Yang-Mills
  amplitudes}},  {\em Nucl. Phys. B} {\bf 913} (2016) 151--162,
  [\href{http://arxiv.org/abs/1606.09616}{{\tt arXiv:1606.09616}}].

\bibitem{Nandan:2016pya}
D.~Nandan, J.~Plefka, O.~Schlotterer, and C.~Wen, {\it {Einstein-Yang-Mills
  from pure Yang-Mills amplitudes}},  {\em JHEP} {\bf 10} (2016) 070,
  [\href{http://arxiv.org/abs/1607.05701}{{\tt arXiv:1607.05701}}].

\bibitem{Schlotterer:2016cxa}
O.~Schlotterer, {\it {Amplitude relations in heterotic string theory and
  Einstein-Yang-Mills}},  {\em JHEP} {\bf 11} (2016) 074,
  [\href{http://arxiv.org/abs/1608.00130}{{\tt arXiv:1608.00130}}].

\bibitem{Fu:2017uzt}
C.-H. Fu, Y.-J. Du, R.~Huang, and B.~Feng, {\it {Expansion of
  Einstein-Yang-Mills Amplitude}},  {\em JHEP} {\bf 09} (2017) 021,
  [\href{http://arxiv.org/abs/1702.08158}{{\tt arXiv:1702.08158}}].

\bibitem{Chiodaroli:2017ngp}
M.~Chiodaroli, M.~Gunaydin, H.~Johansson, and R.~Roiban, {\it {Explicit
  Formulae for Yang-Mills-Einstein Amplitudes from the Double Copy}},  {\em
  JHEP} {\bf 07} (2017) 002, [\href{http://arxiv.org/abs/1703.00421}{{\tt
  arXiv:1703.00421}}].

\bibitem{Teng:2017tbo}
F.~Teng and B.~Feng, {\it {Expanding Einstein-Yang-Mills by Yang-Mills in CHY
  frame}},  {\em JHEP} {\bf 05} (2017) 075,
  [\href{http://arxiv.org/abs/1703.01269}{{\tt arXiv:1703.01269}}].

\bibitem{Du:2017gnh}
Y.-J. Du, B.~Feng, and F.~Teng, {\it {Expansion of All Multitrace Tree Level
  EYM Amplitudes}},  {\em JHEP} {\bf 12} (2017) 038,
  [\href{http://arxiv.org/abs/1708.04514}{{\tt arXiv:1708.04514}}].

\bibitem{Du:2017kpo}
Y.-J. Du and F.~Teng, {\it {BCJ numerators from reduced Pfaffian}},  {\em JHEP}
  {\bf 04} (2017) 033, [\href{http://arxiv.org/abs/1703.05717}{{\tt
  arXiv:1703.05717}}].

\bibitem{Bern:2008qj}
Z.~Bern, J.~J.~M. Carrasco, and H.~Johansson, {\it {New Relations for
  Gauge-Theory Amplitudes}},  {\em Phys. Rev. D} {\bf 78} (2008) 085011,
  [\href{http://arxiv.org/abs/0805.3993}{{\tt arXiv:0805.3993}}].

\bibitem{Bern:2010ue}
Z.~Bern, J.~J.~M. Carrasco, and H.~Johansson, {\it {Perturbative Quantum
  Gravity as a Double Copy of Gauge Theory}},  {\em Phys. Rev. Lett.} {\bf 105}
  (2010) 061602, [\href{http://arxiv.org/abs/1004.0476}{{\tt
  arXiv:1004.0476}}].

\bibitem{DelDuca:1999rs}
V.~Del~Duca, L.~J. Dixon, and F.~Maltoni, {\it New color decompositions for
  gauge amplitudes at tree and loop level},  {\em Nucl. Phys. B} {\bf 571}
  (2000) 51--70, [\href{http://arxiv.org/abs/hep-ph/9910563}{{\tt
  hep-ph/9910563}}].

\bibitem{Hou:2018bwm}
L.~Hou and Y.-J. Du, {\it {A graphic approach to gauge invariance induced
  identity}},  {\em JHEP} {\bf 05} (2019) 012,
  [\href{http://arxiv.org/abs/1811.12653}{{\tt arXiv:1811.12653}}].

\bibitem{Du:2019vzf}
Y.-J. Du and L.~Hou, {\it {A graphic approach to identities induced from
  multi-trace Einstein-Yang-Mills amplitudes}},  {\em JHEP} {\bf 05} (2020)
  008, [\href{http://arxiv.org/abs/1910.04014}{{\tt arXiv:1910.04014}}].

\bibitem{Tian:2021dzf}
H.~Tian, E.~Gong, C.~Xie, and Y.-J. Du, {\it {Evaluating EYM amplitudes in four
  dimensions by refined graphic expansion}},  {\em JHEP} {\bf 04} (2021) 150,
  [\href{http://arxiv.org/abs/2101.02962}{{\tt arXiv:2101.02962}}].

\bibitem{Bern:2010yg}
Z.~Bern, T.~Dennen, Y.-t. Huang, and M.~Kiermaier, {\it {Gravity as the Square
  of Gauge Theory}},  {\em Phys. Rev.} {\bf D82} (2010) 065003,
  [\href{http://arxiv.org/abs/1004.0693}{{\tt arXiv:1004.0693}}].

\bibitem{BjerrumBohr:2010hn}
N.~E.~J. Bjerrum-Bohr, P.~H. Damgaard, T.~Sondergaard, and P.~Vanhove, {\it
  {The Momentum Kernel of Gauge and Gravity Theories}},  {\em JHEP} {\bf 01}
  (2011) 001, [\href{http://arxiv.org/abs/1010.3933}{{\tt arXiv:1010.3933}}].

\bibitem{Du:2011js}
Y.-J. Du, B.~Feng, and C.-H. Fu, {\it {BCJ Relation of Color Scalar Theory and
  KLT Relation of Gauge Theory}},  {\em JHEP} {\bf 08} (2011) 129,
  [\href{http://arxiv.org/abs/1105.3503}{{\tt arXiv:1105.3503}}].

\bibitem{Broedel:2013tta}
J.~Broedel, O.~Schlotterer, and S.~Stieberger, {\it {Polylogarithms, Multiple
  Zeta Values and Superstring Amplitudes}},  {\em Fortsch. Phys.} {\bf 61}
  (2013) 812--870, [\href{http://arxiv.org/abs/1304.7267}{{\tt
  arXiv:1304.7267}}].

\bibitem{Cachazo:2013gna}
F.~Cachazo, S.~He, and E.~Y. Yuan, {\it {Scattering equations and
  Kawai-Lewellen-Tye orthogonality}},  {\em Phys. Rev.} {\bf D90} (2014), no.~6
  065001, [\href{http://arxiv.org/abs/1306.6575}{{\tt arXiv:1306.6575}}].

\bibitem{Cachazo:2013hca}
F.~Cachazo, S.~He, and E.~Y. Yuan, {\it {Scattering of Massless Particles in
  Arbitrary Dimensions}},  {\em Phys. Rev. Lett.} {\bf 113} (2014), no.~17
  171601, [\href{http://arxiv.org/abs/1307.2199}{{\tt arXiv:1307.2199}}].

\bibitem{Cachazo:2013iea}
F.~Cachazo, S.~He, and E.~Y. Yuan, {\it {Scattering of Massless Particles:
  Scalars, Gluons and Gravitons}},  {\em JHEP} {\bf 07} (2014) 033,
  [\href{http://arxiv.org/abs/1309.0885}{{\tt arXiv:1309.0885}}].

\bibitem{Cachazo:2014nsa}
F.~Cachazo, S.~He, and E.~Y. Yuan, {\it {Einstein-Yang-Mills Scattering
  Amplitudes From Scattering Equations}},  {\em JHEP} {\bf 01} (2015) 121,
  [\href{http://arxiv.org/abs/1409.8256}{{\tt arXiv:1409.8256}}].

\bibitem{Mafra:2016ltu}
C.~R. Mafra, {\it {Berends-Giele recursion for double-color-ordered
  amplitudes}},  {\em JHEP} {\bf 07} (2016) 080,
  [\href{http://arxiv.org/abs/1603.09731}{{\tt arXiv:1603.09731}}].

\bibitem{Mafra:2015syt}
C.~R. Mafra and O.~Schlotterer, {\it {A solution to the nonlinear field
  equations of ten dimensional supersymmetric Yang-Mills theory}},  {\em Phys
  Rev.D} {\bf 92} (2015) 066001, [\href{http://arxiv.org/abs/1501.05562}{{\tt
  arXiv:1501.05562}}].

\bibitem{Lee:2016tbd}
S.~Lee, C.~R. Mafra, and O.~Schlotterer, {\it {Non-linear gauge transformations
  in $D = 10$ SYM theory and the BCJ duality}},  {\em JHEP} {\bf 03} (2016)
  090, [\href{http://arxiv.org/abs/1510.08843}{{\tt arXiv:1510.08843}}].

\bibitem{Bridges:2019siz}
E.~Bridges and C.~R. Mafra, {\it {Algorithmic construction of SYM multiparticle
  superfields in the BCJ gauge}},  {\em JHEP} {\bf 10} (2019) 022,
  [\href{http://arxiv.org/abs/1906.12252}{{\tt arXiv:1906.12252}}].

\bibitem{Frost:2020bmk}
H.~Frost, C.~R. Mafra, and L.~Mason, {\it {A Lie bracket for the momentum
  kernel}},  \href{http://arxiv.org/abs/2012.00519}{{\tt arXiv:2012.00519}}.

\bibitem{Wu:2021bcy}
K.~Wu and Y.-J. Du, {\it {Off-shell extended graphic rule and the expansion of
  Berends-Giele currents in Yang-Mills theory}},  {\em JHEP} {\bf 01} (2021)
  162, [\href{http://arxiv.org/abs/2109.14462}{{\tt arXiv:2109.14462}}].

\bibitem{Berends:1987me}
F.~A. Berends and W.~T. Giele, {\it {Recursive Calculations for Processes with
  n Gluons}},  {\em Nucl. Phys. B} {\bf 306} (1988) 759.

\bibitem{BjerrumBohr:2009rd}
N.~E.~J. Bjerrum-Bohr, P.~H. Damgaard, and P.~Vanhove, {\it {Minimal Basis for
  Gauge Theory Amplitudes}},  {\em Phys. Rev. Lett.} {\bf 103} (2009) 161602,
  [\href{http://arxiv.org/abs/0907.1425}{{\tt arXiv:0907.1425}}].

\bibitem{Stieberger:2009hq}
S.~Stieberger, {\it {Open \& Closed vs. Pure Open String Disk Amplitudes}},
  \href{http://arxiv.org/abs/0907.2211}{{\tt arXiv:0907.2211}}.

\bibitem{Chen:2011jxa}
Y.-X. Chen, Y.-J. Du, and B.~Feng, {\it {A Proof of the Explicit Minimal-basis
  Expansion of Tree Amplitudes in Gauge Field Theory}},  {\em JHEP} {\bf 02}
  (2011) 112, [\href{http://arxiv.org/abs/1101.0009}{{\tt arXiv:1101.0009}}].

\bibitem{Kleiss:1988ne}
R.~Kleiss and H.~Kuijf, {\it {Multi - Gluon Cross-sections and Five Jet
  Production at Hadron Colliders}},  {\em Nucl. Phys. B} {\bf 312} (1989)
  616--644.

\bibitem{Feng:2010my}
B.~Feng, R.~Huang, and Y.~Jia, {\it {Gauge Amplitude Identities by On-shell
  Recursion Relation in S-matrix Program}},  {\em Phys. Lett.} {\bf B695}
  (2011) 350--353, [\href{http://arxiv.org/abs/1004.3417}{{\tt
  arXiv:1004.3417}}].

\bibitem{Ma:2011um}
Q.~Ma, Y.-J. Du, and Y.-X. Chen, {\it {On Primary Relations at Tree-level in
  String Theory and Field Theory}},  {\em JHEP} {\bf 02} (2012) 061,
  [\href{http://arxiv.org/abs/1109.0685}{{\tt arXiv:1109.0685}}].

\bibitem{Frost:2021asr}
H.~Frost, {\it {The Algebraic Structure of the KLT Relations for Gauge and
  Gravity Tree Amplitudes}},  {\em SIGMA} {\bf 17} (2021) 101,
  [\href{http://arxiv.org/abs/2111.07257}{{\tt arXiv:2111.07257}}].

\end{thebibliography}\endgroup

\end{document}